\documentclass[authoryear]{elsarticle}
\usepackage{lineno}

\usepackage{natbib}
\usepackage{amssymb}
\usepackage{amsmath}
\usepackage{color}
\usepackage{url}
\usepackage{booktabs}
\usepackage{subcaption}
\usepackage[]{units}
\usepackage{siunitx}
\usepackage{multirow}

\journal{Renewable Energy}

\newcommand{\dx}{\,\mathrm{d}\vec x}
\newcommand{\dt}{\,\mathrm{d}t}

\newcommand{\Omegafarm}{\Omega_{\textrm{farm}}}
\newcommand{\dtmax}{\overline{d}}
\definecolor{dark_grey}{gray}{0.3}

\newtheorem{algorithm}{Algorithm}

\newcommand{\uvec}{{\vec u}}

\newcommand{\nvec}{{\vec n}}
\newcommand{\xvec}{{\vec x}}
\begin{document}

\begin{frontmatter}

\title{Design optimisation and resource assessment for tidal-stream renewable
energy farms using a new continuous turbine approach
}

\author[simula]{S.W. Funke\corref{cor1}}
\ead{simon@simula.no}
\author[ic]{S.C. Kramer}
\author[ic,grant]{M.D. Piggott}

\address[simula]{Center for Biomedical Computing, Simula Research Laboratory, Oslo, Norway}
\address[ic]{Applied Modelling and Computation Group, Department of Earth Science and Engineering, Imperial College London, London, UK}
\address[grant]{Grantham Institute for Climate Change, Imperial College London, London, UK}

\begin{abstract}
    This paper presents a new approach for optimising the design of tidal stream
    turbine farms.  In this approach, the turbine farm is represented by a
    turbine density function that specifies the number of turbines per unit
    area and an associated continuous locally-enhanced bottom friction field.
    The farm design question is formulated as a mathematical optimisation
    problem constrained by the shallow water equations and solved with
    efficient, gradient-based optimisation methods.
    The resulting method is accurate, computationally efficient, allows
    complex installation constraints, and supports different goal quantities
    such as to maximise power or profit. The outputs of the optimisation
    are the optimal number of turbines, their location within the
    farm, the overall farm profit, the farm's power extraction, and the installation cost.

    We demonstrate the capabilities of the method on a validated numerical
    model of the Pentland Firth, Scotland. We optimise the design of four tidal farms
    simultaneously, as well as individually, and study how farms in close proximity
    may impact upon one another.


\end{abstract}


\end{frontmatter}

\section{Introduction}
Tidal stream turbines represent one of the most promising technologies for capturing
marine energy. These turbines are installed in fast tidally-induced currents,
where they convert mechanical energy into electricity in a similar way to wind
turbines. Farms comprising hundreds of turbines have the potential to produce hundreds
of megawatts of highly predictable power.  However, in order to be economically
viable, these farms must be carefully designed. Two of the most critical
design decisions are (i) the locations of the turbines, and (ii) the number of
turbines making up the farm.

Finding good turbine locations is
challenging for a number of reasons. Firstly, various practical constraints need to be
taken into account that pose limitations on where turbines may be installed, for
instance turbines may not be installed in areas which are too steep, shallow or deep.
Secondly, a tidal turbine has both local effects on the flow (for instance
reduced flow speed downstream of the turbine due to wake effects, and accelerated flow
to the sides of the swept area), and basin-wide effects (reduced flow speeds due
to large-scale blockage effects). Furthermore, the turbine's energy output depends cubically
on the flow speed. Combining these effects means that the performance of each
turbine depends on the position of all other turbines, and that even small changes in
a turbine's location may result in a significant change in the total farm
performance \citep{sutherland2007, polagye2011, draper2013, funke2014}. At larger geographical scales, the design of an
individual array may have implications for the optimal location, size and design
of other arrays in the vicinity.
It is therefore potentially important to consider both inter- and intra- array
design and competition issues in a fully coupled manner.

The optimal number of turbines in a farm is the result of multiple competing factors.
Adding turbines generally (up to a point) increases the farm's total power production, but also increases the
total project cost. Furthermore, the energy available to a farm should be
considered as a limited resource, which means
that the
power production per turbine decreases with increasing numbers of
turbines \citep{vennell2015}. As a result, the overall profitability of the farm
reduces if too few or too many turbines are installed in the farm.

When combining all these factors it becomes clear that determining the best farm design
is a complex, tightly coupled optimisation problem.  The simplest approach to tackle
this problem is to select some potential designs and to use a
hydrodynamic model to predict their
performance~\citep{lee2010,ahmadian2012,divett2013}. However, in practice the number of
possible designs is too large to be fully explored.  Another option is to
simplify the hydrodynamic model such that an analytical solution can be
derived. Then all possible farm designs can be analysed and the optimum design
derived directly~\citep{garrett2005,vennell2010,vennell2011}. A more recent
approach uses an accurate hydrodynamic model and applies efficient, derivative-based optimisation methods to
determine optimal farm designs~\citep{funke2014,culley2015,barnett2014}.

This paper presents a new \emph{continuous turbine approach} for designing tidal
turbine farms and for resource assessment. The basic idea is similar to
\cite{funke2014} in that we formulate the farm design problem as a mathematical
optimisation problem, describe the hydrodynamics by the shallow water
equations, and solve the problem with derivative-based optimisation methods. The novelty of this work is that the tidal
farm is represented as a turbine density function, in contrast to resolving individual turbines as in
\cite{funke2014}. This turbine density function has high values in areas with
many tidal turbines and vanishes in areas with no turbines.

The proposed approach offers a number of key benefits. It is \emph{accurate},
because the interaction between farm and flow is treated in a fully coupled manner, here through
the solution of the shallow water equations.  It is \emph{flexible} in that it naturally
supports minimum distance constraints between turbines,  feasible turbine
locations can be chosen arbitrarily, and the goal quantity defining the
optimisation problem can be a
combination of farm profit, installation cost, and other user-defined
quantities. Furthermore, the approach is \emph{efficient} because the total
number as well as the positions (or rather local density) of the turbines are simultaneously optimised for, because the
numerical mesh does not need to resolve individual turbines, and because the
derivative-based optimisation methods scale well with the farm size. In fact,
from our experience computational cost is nearly independent
on the farm size and the number of turbines.
As a result, this approach is
suitable for designing large farms with hundreds of turbines and complex
constraints, and further allows for the design of multiple (potentially interacting and competing) farms simultaneously.

More specifically, the novel contributions of this work are the following:
\begin{itemize}
    \item We formulate the turbine farm design problem and resource
        assessment in terms of a mathematical optimisation problem. The farm is
        represented by a turbine density function (sections
        \ref{sec:problem_formulation}--\ref{sec:form_opt_problem}). The goal
        is to identify the farm configuration which maximises
        the farm's financial profit.
    \item We show how to discretise,
        and efficiently solve this optimisation problem using derivative-based
        optimisation algorithms and the adjoint equations~(section \ref{sec:numericalsolution}). The
        implementation is available as part of the open-source software
        OpenTidalFarm (\url{opentidalfarm.org}).
    \item  We show that the results from the new continuous approach are consistent with
    those obtained using the approach presented in
        \citet{funke2014}, where turbines were resolved explicitly (section
        \ref{sec:contvsdisc}).
    \item We validate the underlying hydrodynamic model against observations made in
    the Pentland Firth (section \ref{sec:pentland}).
    \item We demonstrate the efficiency of the approach by optimising four turbine
    farms in the Pentland Firth,
        first individually and then simultaneously,
        and compare the farm designs (section
        \ref{sec:pentland-firth-optimisation}). We show
        that farms have the potential to significantly impact one another, and that
        this interaction should be taken into account during the design
        process.
        To our knowledge this is the first example of a simultaneous multi-farm design optimisation.
\end{itemize}

\section{Problem formulation}\label{sec:problem_formulation}
\subsection{Farm optimisation}
\begin{figure}[t]
    \centering
    \includegraphics[width=0.45\textwidth]{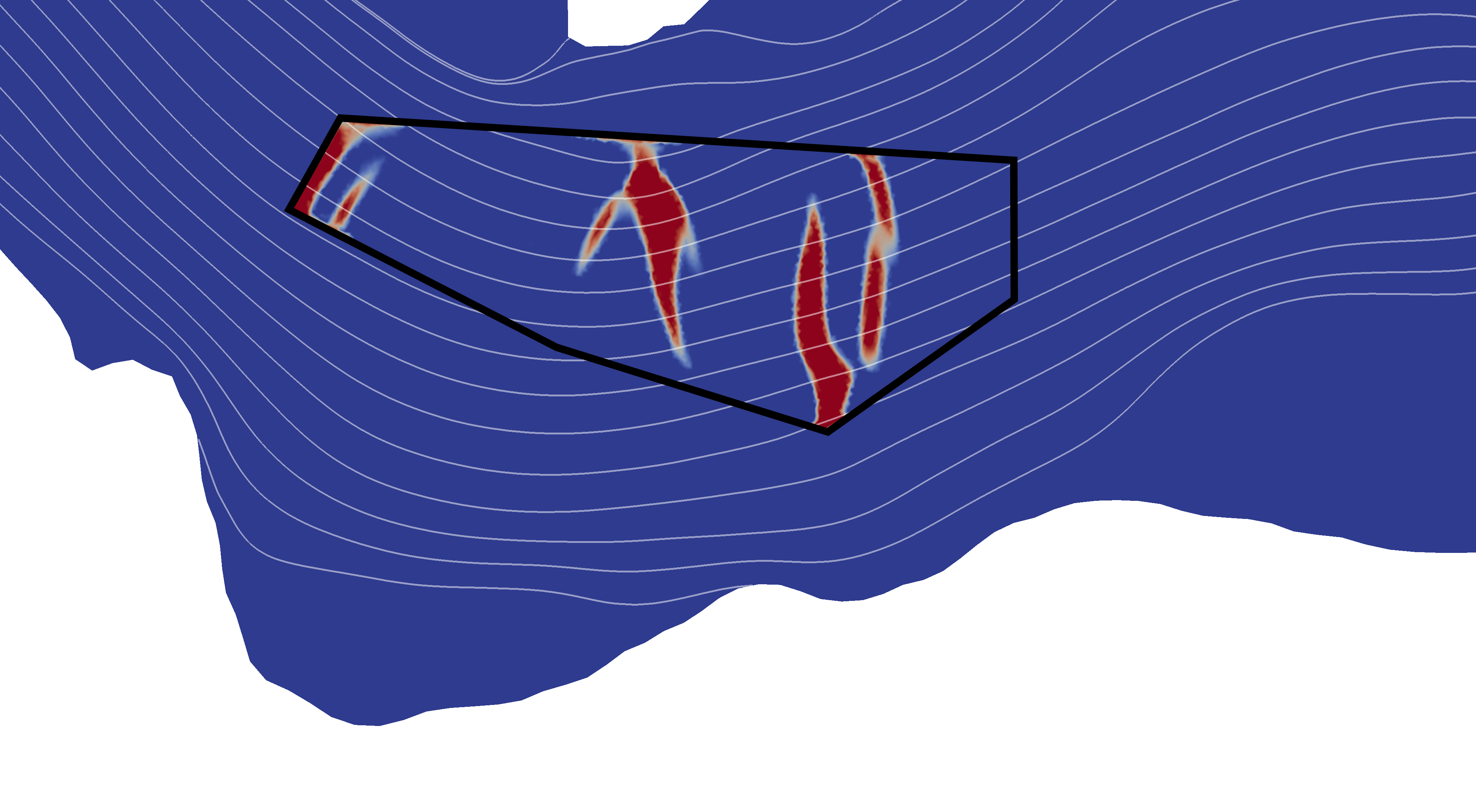}
    \includegraphics[width=0.45\textwidth]{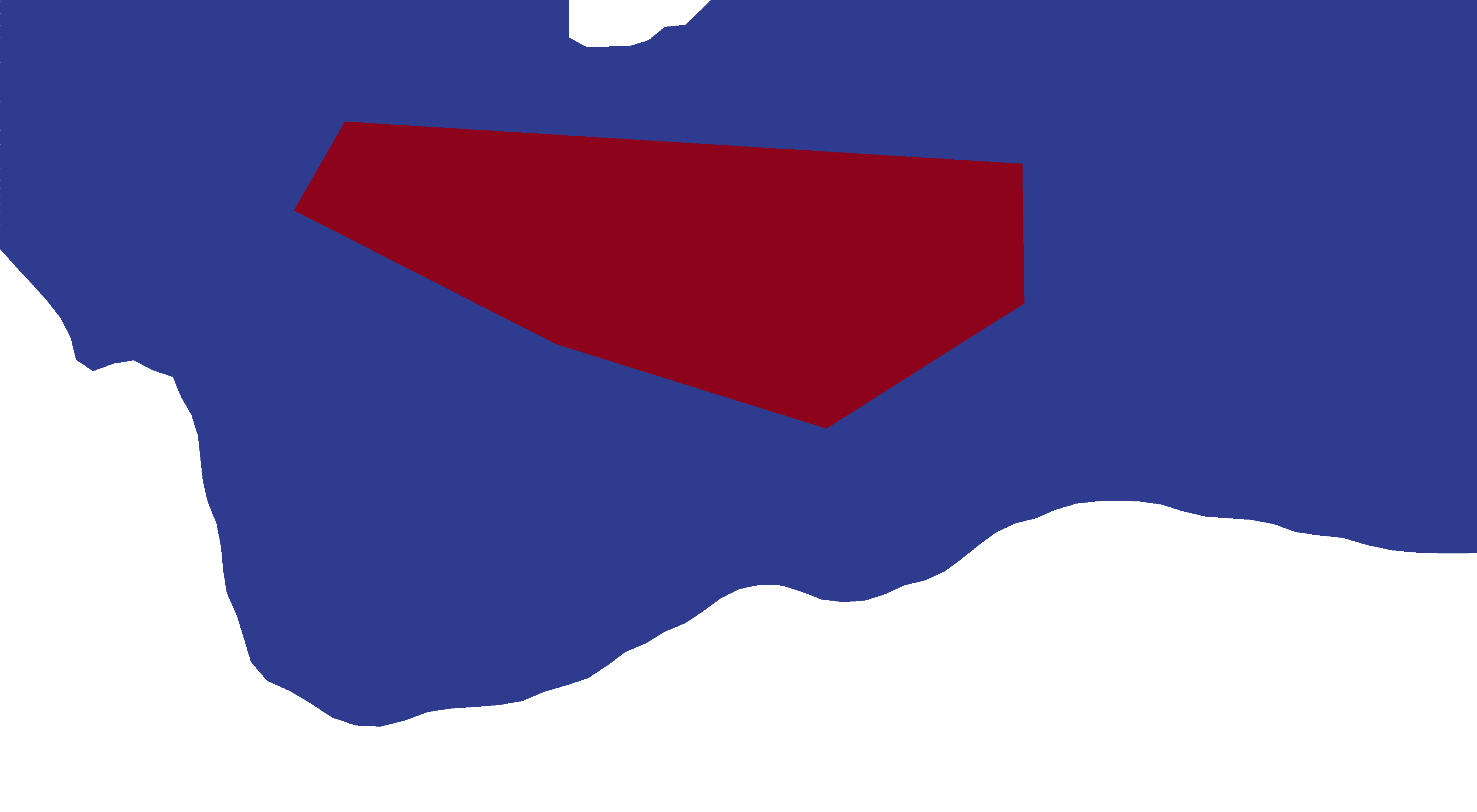}
    \caption{Solution of the tidal farm design problem \eqref{eq:max_problem}
        for a site in the Inner Sound of the
        Pentland Firth. The image on the left shows the optimal turbine density field
        $d(x)$ with flow streamlines and farm boundary indicated. The image on the right
        shows the upper density bound $\bar d$, which here takes the value zero outside the
        array area.
        The optimisation predicts an optimal number of $266$
        turbines, and a farm power production of $316$ MW.
         Only a subset of the entire computational domain is shown here, refer to section \ref{sec:pentland-firth-optimisation} for further details.}
\label{fig:example_optimal_friction}
\end{figure}
We formulate the farm design (the total number and positions of turbines) question via
the optimisation problem:

\begin{equation}
\begin{aligned}
    & \max_{d} \; \textrm{Profit}(d) \\
    & \textrm{subject to }  \; 0 \le d \le \bar d.
\end{aligned}
    \label{eq:max_problem}
\end{equation}

\noindent Here, `Profit$(\cdot)$' is a function which evaluates the financial profit of
the farm over its entire lifetime and is the goal quantity to be maximised. The farm design is described by a spatially
varying turbine density function $d: \Omega \to \mathbb R$ (e.g. figure
\ref{fig:example_optimal_friction}, left), where $\Omega
\subset \mathbb R^2$ is the horizontal domain. That is, an area with $d(\xvec)=0$ has
no turbines, and an area with $d(\xvec)=\bar d(\xvec)$ has densely deployed turbines. The upper bound $\bar d$ is defined by the user and specifies
the maximum number of turbines per unit area: it is zero where no turbines may
be installed (outside the lease area, or in too steep, shallow or deep areas as shall be demonstrated in section \ref{sec:pentland-firth-optimisation}),
and a positive constant elsewhere (e.g. figure \ref{fig:example_optimal_friction}, right).  This constant could be derived from a dense grid
layout: for a minimal distance $D_\textrm{min}$ between turbines it is $\bar d(\xvec)
= 1/{D_{\textrm{min}}^2}$. This could trivially be extended to account for
different inter- and intra- row spacing constraints.

The solution of the farm design problem \eqref{eq:max_problem} yields several useful insights.
Firstly, the optimal turbine density field $d(x)$ indicates where the turbines should be
installed. Secondly, the optimal number of turbines $N$ is obtained by integrating the
optimal density function, that is
\begin{equation}
    N = \int_\Omega d(\xvec) \dx.
    \label{eq:opt_number_turbine}
\end{equation}
Finally, the Profit function provides a
good indicator for the potential of the tidal resource for a given farm site.

\subsection{Profit model}
A profit model for a tidal farm can be very complex and include various factors such as the
farm design, turbine properties, installation costs, learning rates and the underlying energy
market \citep{roc2015}. Here, we consider for simplicity only
two dominant components:  the revenue of
selling the produced energy and the cost of buying, installing and
maintaining the turbines, that is:
\begin{equation}
    \textrm{Profit}(d) = \textrm{Revenue}(d) - \textrm{Cost}(d).
    \label{eq:profit_model}
\end{equation}

The revenue of the tidal farm over its lifetime is modelled here as
\begin{equation}
    \textrm{Revenue}(d) = I k E(\uvec, d),
    \label{eq:revenue_model}
\end{equation}
where $I$ denotes the financial income per energy unit, $0 \le
k \le 1$ is a turbine efficiency coefficient that takes mechanical and
electrical losses into account, and $E(\uvec, d)$ is the total amount of energy that the farm
(encoded by $d$) extracts from the flow velocity $\uvec$ over its lifetime. Note that
$\uvec$ can itself be considered a function of $d$.

A very simple cost model is employed here that scales linearly with the number of installed turbines:
\begin{equation*}
    \textrm{Cost}(d) = C \int_{\Omega} d(\xvec) \dx,
 \label{eq:cost_model}
\end{equation*}
where $C$ is the financial cost of a single turbine. Given \eqref{eq:opt_number_turbine} this of course is equivalent to
$\textrm{Cost}=CN$.
In the special case of setting the turbine cost to zero, the solution to
\eqref{eq:max_problem} yields the maximum extractable energy from the
farm site.

Note that the Revenue and Cost models do not take a discount
rate into account, and no distinction is made between upfront investments and
e.g. maintenance costs in the Cost model. Although these are essential factors to obtain a
realistic estimate of the profitability of a farm, these have been left out
here for simplicity but could easily be incorporated when realistic values are
known.

\section{The energetics of a tidal turbine farm}
\subsection{Shallow water equations}
The farm energy extraction $E(\uvec, d)$ depends on the flow velocity $\uvec$, which
in turn depends on the farm design $d$.
Here we model these dynamics using the time-dependent,
nonlinear shallow water equations with the presence of turbines
represented through a locally-increased bottom
friction coefficient. The resulting equations are
\begin{subequations}
\begin{equation}
 \frac{\partial  \uvec}{\partial t} +  \uvec \cdot \nabla  \uvec - \nu \nabla^2
 \uvec  + g \nabla \eta + \frac{c_b + c_t(d)}{H} \| \uvec\|  \uvec = 0, \\
 \label{eq:sw_momentum}
\end{equation}
\begin{equation}
 \frac{\partial \eta}{\partial t} + \nabla \cdot \left(H \uvec\right) = 0, \\
 \label{eq:sw_continuity}
\end{equation}
\label{eq:sw_equations}%
\end{subequations}
with appropriate initial and boundary conditions.
Here $\uvec: \Omega \times (0, T) \to \mathbb R^2$ is the depth-averaged
velocity, $\eta: \Omega \times (0, T) \to \mathbb R$ is the free-surface
displacement, $H: \Omega \to \mathbb R$ is the total water depth (that is
$H=\eta + h$, where $h$ is the water depth at rest), $c_b:
\Omega \to \mathbb R$ is the natural background friction which here is assumed
to be constant, and $c_t(d): \Omega \to
\mathbb R$ is the additional farm induced friction field that represents the drag exerted by the
turbine farm on the flow. The viscosity term, with kinematic viscosity $\nu:
\Omega\to\mathbb R$, represents mixing through sub-grid scale processes. A fixed
eddy viscosity has been chosen in the runs presented here.

\subsection{Relationship between turbine density and friction}

To derive the relationship between the farm induced friction
$c_t$ and the turbine density function $d$, we use an idea very similar to the
enhanced bottom drag formulations \citep{divett2013,funke2014,martin2015}, where the
turbine induced drag is chosen such that the resulting force approximates the drag
force of an analytical model of the turbines.

To calculate the force associated with the farm induced friction $c_t$,
we consider the depth-integrated momentum
equation which can be derived from the shallow water equations
\eqref{eq:sw_equations}:
\begin{equation*}
  \rho \left( \frac{\partial H \uvec}{\partial t} +
  \nabla \cdot \left(H \uvec\otimes\uvec\right)
  - \nu H \nabla^2  \uvec  + gH \nabla \eta\right) +
  \rho(c_b + c_t(d)) \| \uvec\|  \uvec = 0.
 \label{eq:sw_momentum_force}
\end{equation*}

When integrated over some arbitrary area, the first term represents the change of momentum
in that area. The integration of the subsequent terms gives sources and sinks
in this momentum balance, for instance the integrated advection term can be written in terms of momentum fluxes
into and out of the area. The integral of the friction term gives a momentum sink.
Thus, the force produced by the farm induced friction $c_t$ is given by:
\begin{equation}
    \textbf{F}_{\textrm{farm}}(t) = \rho \int_\Omega c_t(d(\xvec)) \| \uvec(\xvec, t) \|
    \uvec(\xvec, t) \dx. \label{eq:force_farm}
\end{equation}

We now consider a single turbine in a three-dimensional flow with upstream velocity
$\uvec_{\infty}$. Its drag force $\textbf{F}_\textrm{turbine} \in \mathbb R^3$
is commonly parameterised as
\begin{equation*}
\textbf{F}_{\textrm{turbine}}(\uvec_\infty) = \frac{1}{2} \rho C_t A_t
  \|\uvec_{\infty}\|\uvec_{\infty},
 \label{eq:force_produced_by_3D_turbine}
\end{equation*}
where $C_t$ is the (dimensionless) drag coefficient of the turbine, $A_t$ is
the surface area swept by the turbine's blades, and $\uvec_\infty$ is the
free-stream velocity. The drag of a farm made up of $N$ devices is therefore given by:
\begin{equation*}
  \textbf{F}_{\textrm{farm}} = \sum_{i=1}^N \frac{1}{2} \rho C_t A_t
  \|\uvec_{\infty,i}\|\uvec_{\infty,i}.
\end{equation*}
Here we assume that every turbine has the same thrust coefficient $C_t$ and
cross-sectional surface area $A_t$ but take into account a different free-stream
velocity $\uvec_{\infty,i}$. The interpretation of free-stream velocity in the
case of multiple inter-dependent turbines is not straight-forward. For resolved
three-dimensional models this issue can be circumvented by rewriting the force
as a function of the turbine velocity \citep{roc2015}. In depth-averaged models
however, the local velocity available in the model
will be somewhere between the free-stream and turbine velocities, based both on vertical considerations as well as the fact that the less
well-resolved the model is in the horizontal (and therefore the more the friction is `spread out' over the computational mesh) the
closer the local model velocity is to the idealised free-stream velocity (see
\citet{kramer2015} for a more detailed discussion). In the turbine density
approach proposed here, individual turbines are not in any sense individually
resolved and therefore the
local model velocity at the turbine location may be a good approximation to the free-stream velocity.

Next, we turn
the farm drag force from a sum over the discrete turbines into an integral
of continuous drag scaled by the turbine density function $d$:
\begin{equation}
 \textbf{F}_{\textrm{farm}}(t) =  \int_\Omega \frac{1}{2} \rho C_T A_t
 d(\xvec)\|\uvec{(\xvec, t)}\|\uvec{(\xvec, t)}\dx.
    \label{eq:farm_force_from_turbine_density}
\end{equation}
Comparing \eqref{eq:force_farm} with \eqref{eq:farm_force_from_turbine_density},
we see that to obtain the desired farm drag, we need to set
\begin{equation}
    c_t(d(\xvec)) = \frac 12 C_T A_T d(\xvec).
  \label{eq:drag_from_density}
\end{equation}

\subsection{Farm production}
The farm's energy extraction, $E$, in \eqref{eq:revenue_model} is
computed from the hydrodynamic model.
Taking the dot product of \eqref{eq:force_farm} with $\uvec$ yields the farm's power extraction from the flow:
\begin{equation}
    P_{\text{farm}}(t) = \rho \int_\Omega c_t(d(\xvec)) \| \uvec(\xvec, t) \|^3 \dx.
 \label{eq:power_extracted}
\end{equation}
The total energy extraction is obtained by integrating the power over the farm's
lifetime $L$:
\begin{equation*}
    E(\uvec, d) = \rho \int_{0}^{L}\int_\Omega c_t(d(\xvec)) \| \uvec(\xvec, t) \|^3 \dx \dt.
\end{equation*}
It should be noted that the energy extracted from the flow by the enhanced
bottom drag term will generally be an over-estimate of the energy that can
be directly taken out of the flow by the turbines; part of the energy loss is
due to mixing losses (e.g. vertical mixing) that are not explicitly represented
in the model \citep{kramer2015}. To compensate for this in the Revenue model, the percentage of
usefully extractable energy can be incorporated in the factor $k$ in
\eqref{eq:revenue_model}.

\subsection{Blockage effects and rating}
\label{blockage_and_rating}
For a free standing turbine
the theoretical maximum for the thrust coefficient is given by the
Lanchester-Betz limit: $C_T=8/9$. Depending on the ratio of the turbine cross-section with
respect to the cross-section of the farm over its entire width (local blockage)
and with respect to the total
channel cross section (global blockage) however, this limit can be
considerably higher. These blockage effects are not correctly incorporated
in the depth-averaged approach with constant $C_T$ described here because
the drag is effectively applied over the entire water depth, and in particular
local blockage effects cannot be explicitly modelled. Several papers have derived
analytical expressions (see e.g. \citet{garrett2008, whelan2009, nishino2012})
that calculate blockage effects in idealized circumstances.
It is a matter of ongoing research to investigate whether these expressions can be
incorporated within the depth-averaged drag approach to correct for blockage
effects. See \citet{vogel2014} for an example of a blockage correction applied to
a tidal fence in a depth-averaged model.

Another effect that is not taken into account by using a constant $C_T$ is the
fact that turbines are typically tuned to maintain a constant power output above
a certain rated speed, by dropping the effective thrust coefficient with
increasing speed. Thus instead of a constant $C_T$, a thrust curve should be
considered, i.e. $C_T(\vec u)$ is a function of the flow speed $\vec u$, and the thrust force is no
longer quadratic. Exact details for turbines that are expected to be installed
in the sites that are investigated in this paper are hard to obtain. For
simplicity we therefore follow the constant $C_T$ approach described above.
It should be noted however that when these details are known, the optimisation
approach described in this paper can be easily extended to use generic
thrust curves. Likewise if algebraic expressions can be derived to incorporate
blockage effects more accurately --- this is a matter of further research --- such
expressions can also easily be incorporated in the optimisation procedure.

\section{Formulation as an optimisation problem}\label{sec:form_opt_problem}
Reconsider the profit model \eqref{eq:profit_model}:
\begin{equation}
    \textrm{Profit}(d) = I k E(\uvec,d) - C \int_\Omega d(\xvec) \dx.
  \label{eq:profit}
\end{equation}
Unfortunately, determining the cost of a turbine $C$, the income factor $I$ and
the turbine efficiency coefficient $k$ requires detailed financial and
technical specification of tidal turbines
and predictions of the electricity market, which are difficult to acquire or estimate.
However, the necessary information can be condensed into a single coefficient by
observing that the solution of problem \eqref{eq:max_problem} is invariant with respect to scaling.
That is we can divide the goal quantity \eqref{eq:profit} by $LIk$ and obtain the
full \textit{continuous farm optimisation problem}:
\begin{equation}
\begin{aligned}
    & \max_{d} \; \frac{1}{L} E(\uvec, d) - \frac{C}{LIk} \int_{\Omega} d(\xvec) \dx \\
 \textrm{subject } & \textrm{to} \\
& \frac{\partial  \uvec}{\partial t} +  \uvec \cdot \nabla  \uvec - \nu
  \nabla^2  \uvec  + g \nabla \eta + \frac{c_b + c_t(d)}{H} \| \uvec\|  \uvec = 0, \\
& \frac{\partial \eta}{\partial t} + \nabla \cdot \left(H \uvec\right) = 0, \\
    & 0 \le d(\xvec) \le \dtmax(\xvec) \qquad \forall \xvec \textrm{ in } \Omegafarm, \\
    & d(\xvec) = 0 \qquad \quad \qquad \forall \xvec \textrm{ in } \Omega \setminus \Omegafarm.
\end{aligned}\label{eq:smeared_optimisation_problem}
\end{equation}
The first two constraint equations are the shallow-water equations \eqref{eq:sw_equations}.
The final two constraints ensure that the turbine density is positive, is
limited by an upper bound in the farm area, and is zero
outside the farm area.

Some remarks can be made about the optimisation problem
\eqref{eq:smeared_optimisation_problem}:
\begin{itemize}
    \item The original goal quantity \eqref{eq:profit_model} computes the profit
      over the farm's lifetime and can be measured in, for instance, US dollars. In
      contrast, the rescaled goal quantity in \eqref{eq:smeared_optimisation_problem}
      is measured in units of power. The rescaled goal quantity should be
      interpreted as the part of the farm's average power extraction which
      contributes to the farm's profit. Hence, if the goal quantity is positive
      the farm produces a positive return, otherwise a negative. In other words,
      $C/(LIk)$ is the average amount of power per turbine that needs to be
      produced in order for the farm to break even.
  \item The optimal turbine density function $d$ can be used to determine the
    best turbine locations in the farm. It can also be used to estimate the
    optimal number of turbines with equation~\eqref{eq:opt_number_turbine}. This is a distinct
    advantage over approaches where turbines are resolved
    individually and the inclusion of the number of turbines as part of the
    optimisation results in a mixed integer problem which are generally
    harder to solve and more costly \citep{culley2014b}.
  \item
    The first term of the goal quantity
    evaluates the farm's average power production over its entire lifetime. In order to
    limit the computational expense this term can be estimated by computing the
    average power production over a smaller time window, for example one tidal
    cycle. To even further reduce the computational expense, a steady state
    solution at a representative time can be used.
  \item Due to the non-negativity constraint placed upon $d$, the goal quantity of
    \eqref{eq:smeared_optimisation_problem} can be reformulated as:
    \begin{equation*}
        \frac{1}{L}\left(E(\uvec, d) - \frac{C}{Ik} \int_\Omega |d(\xvec)| \dx
    \right).
    \end{equation*}
    In other words, the cost term can be identified as a scaled $L^1$-norm of the control function.  In an optimisation context, such a term
    is called $L^1$-regularisation. Generally in optimisation, regularisation
    terms are used to promote optimal solutions with certain properties. For example in this
    case, it is used to promote farm designs with minimal cost. In addition, the
    $L^1$-regularisation has another property: it promotes sparse
    farm designs, that is, designs where the turbine density either vanishes or
    reaches its maximum value. This property is desirable for example to avoid high
    cabling costs \citep{culley2015}. Additionally, it simplifies the
    translation of the optimal turbine density into the actual positioning
    of individual turbines, because areas with maximum density can simply
    be packed with turbines according to the minimum distance constraint.

    From a computational viewpoint, $L^1$-regularisation is
    challenging due to its non-differentiability at $0$ which requires
    special treatment in the optimisation algorithm. However, this problem
    is avoided here thanks to the positivity constraint for $d$, which
    means that standard algorithms can be applied directly.
    \item
        The set of feasible farm designs (that is the set of turbine density functions
        that satisfy the last two inequality constraints in problem
        \eqref{eq:smeared_optimisation_problem}) is convex.  This convexity
        property is an important requirement for gradient-based optimisation
        methods to guarantee convergence. In particular, it is convex for any
        farm geometry (including disconnected and non-convex geometries), turbine constraints and minimum distance
        constraints between turbines. This is an advantage over the
        formulation in \citet{funke2014}, where the farm geometry or
        constraints can result in a non-convex feasible design set.

\end{itemize}

\subsection{Estimation of cost coefficient}
\label{sec:cost-coefficient}
The value ${C}/(LIk)$ in \eqref{eq:smeared_optimisation_problem} can be estimated as follows.
Consider a model turbine with a total energy extraction $E_{\textrm{turbine}}$
from the flow over its lifetime and with a profit margin $m$. The definition of profit margin yields:
\begin{align*}
  m & = \frac{\textrm{Revenue}-\textrm{Cost}}{\textrm{Revenue}}
    = \frac{I k E_{\textrm{turbine}} -  C}{ I k
  E_{\textrm{turbine}}} \nonumber \\
    \Rightarrow  \quad \frac{C}{LIk} & = (1-m) \frac{E_{\textrm{turbine}}}{L}.
\end{align*}
The second factor on the right hand side describes the average power production of the
model turbine over the period $L$.
For a simplified sinusoidal tidal cycle with a peak velocity magnitude
$u_{\textrm{peak}}$ and a peak power production $P_{\textrm{peak}}$, the
average power production
can be estimated as $0.42 P_{\textrm{peak}}$.
Using \eqref{eq:power_extracted}, substituting \eqref{eq:drag_from_density}
for the density function corresponding to a single turbine,
i.e. $\int_\Omega d(\xvec)\dx=1$, we get
\begin{align*}
    P_{\textrm{peak}} = \rho \int_\Omega \tfrac 12 C_t A_t d(\xvec)  u_\textrm{peak}^3\dx =
    \frac 12 C_t A_t \rho u_\textrm{peak}^3.
\end{align*}
This yields the formula
\begin{equation}
  \frac{C}{LIk} = \frac{0.42}2 C_TA_T(1-m) \rho u_{\textrm{peak}}^3.
  \label{eq:cost-coefficient}
\end{equation}

Note that the right hand side only depends indirectly on the turbine efficiency
$k$ and its lifetime $L$, through the estimated profit margin $m$. For a constant
flow (instead of a sinusoidal one), equation \eqref{eq:cost-coefficient} can
be used without the $0.42$ factor.



\section{Numerical solution}\label{sec:numericalsolution}
\begin{figure}
    \centering
    \includegraphics[width=0.8\textwidth]{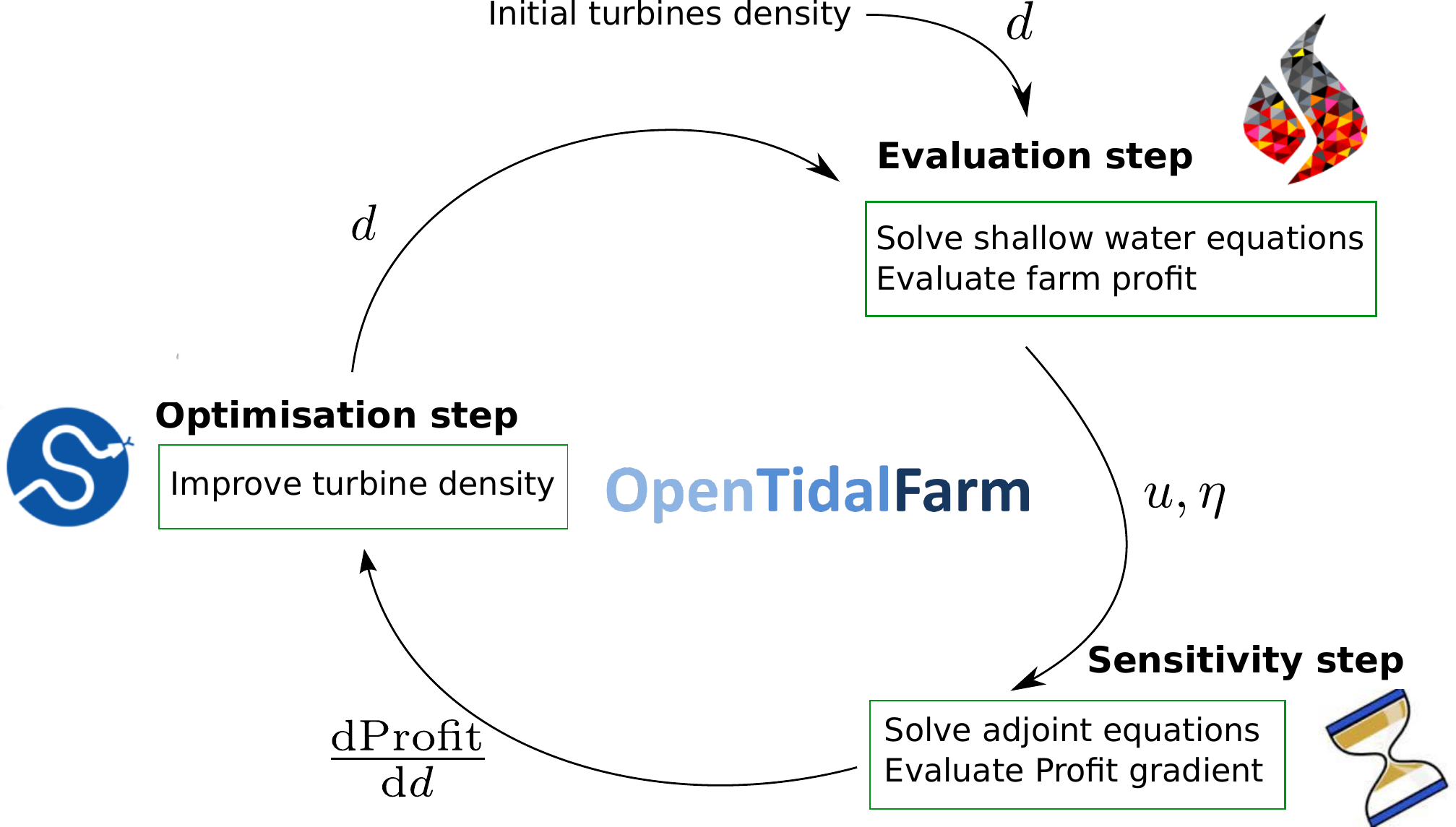}
    \caption{The optimisation loop of OpenTidalFarm and the software it relies
        on.  OpenTidalFarm uses the finite element framework
        FEniCS~\citep{logg2012} to assemble and solve the shallow water and
        adjoint shallow water equations~\citep{funke2014}.  The adjoint shallow
        water equations are automatically derived by dolfin-adjoint and its
        optimisation capabilities~\citep{farrell2013, funke2013}. We employ the
        L-BFGS-B method to solve the optimisation problem~\citep{byrd1995,
        zhu1997b}.
        Each optimisation iteration consists of one (or
        multiple if a line search is required) solution of the shallow water equations
        to evaluate the farm profit, and one solution of the associated adjoint equation to
        compute the discretely consistent profit gradient.
        By combining gradient-based optimisation with the adjoint
        approach, the required optimisation iterations are
        practically independent of the size of the turbine farm.}
\label{fig:software}
\end{figure}

We solve problem~\eqref{eq:smeared_optimisation_problem} by discretising the
shallow water equations with the finite element method and by solving the resulting discrete
optimisation problem with a gradient-based optimisation method. The gradient of the goal quantity, that is the derivative of the profit function with
respect to the turbine density function, is computed with the
adjoint equations.
The iteration
loop of the optimisation is visualised in figure \ref{fig:software}.
The implementation is available as part of OpenTidalFarm
(\url{opentidalfarm.org}), an open-source software package for tidal resource assessment
and turbine farm design optimisation.

\subsection{Spatial and temporal discretisation}\label{sec:discretisation}
We follow \cite{funke2014} to spatially discretise the shallow water equations
with a second-order finite element scheme.
Multiplying the two equations making up the shallow water system by test
functions $\vec \Psi$ and $\Phi$ respectively, integrating over the domain and integrating the
viscosity and divergence terms by parts yields the weak formulation of the
shallow water equations: find $(\uvec, \eta) \in V \times Q$ such that
for all $({\vec\Psi}, \Phi) \in V \times Q$:
\begin{equation}
\begin{split}
    \left\langle \frac{\partial  \uvec}{\partial t},  {\vec\Psi} \right\rangle_\Omega +
    \left\langle  \uvec \cdot \nabla  \uvec,  {\vec\Psi} \right\rangle_\Omega + \nu
    \left\langle\nabla  \uvec, \nabla  {\vec\Psi}\right\rangle_\Omega & \\
    + g \left\langle \nabla \eta,  {\vec\Psi} \right\rangle_\Omega  +
    \left\langle \frac{c_b + c_t(d)}{H} \| \uvec\|  \uvec,  {\vec\Psi}
 \right\rangle_\Omega &=  ~0, \\
\left\langle \frac{\partial \eta}{\partial t}, \Phi \right\rangle_\Omega -
\left\langle H \uvec, \nabla \Phi \right\rangle_\Omega + \left\langle H  \uvec
\cdot  \nvec , \Phi\right\rangle_{\partial \Omega \setminus \partial \Omega_N} &=  ~0.
\end{split}\label{eq:to_spatially_discretised_shallow_water_eq}
\end{equation}
Here $\left\langle \cdot, \cdot \right\rangle$ denotes the
$L^2$ inner product and $\nvec$ is the unit outward facing normal vector on the boundary $\partial \Omega$.
For these equations it is natural to choose
the Hilbert spaces $H^1(\Omega)^2$ and $H^1(\Omega)$ for $V$ and $Q$, respectively.

Dirichlet boundary conditions for the velocity and free surface are imposed by
restricting the function spaces $V$ and $Q$ to functions that satisfy the
boundary conditions, and by using test functions that vanish on the
boundary. The free slip boundary conditions $\uvec\cdot\nvec=0$ on $\partial \Omega_N$ are enforced by
omitting the surface integral on $\partial \Omega_N$.

The discrete problem is obtained by replacing the function spaces with
finite-dimensional subspaces $V_h\subset V$, $Q_h\subset Q$. For their construction one first
creates a triangulation of the domain. $V_h$ is then chosen as continuous,
piecewise quadratic functions and $Q_h$ as the space of continuous,
piecewise linear functions, that is the classical Taylor-Hood ($P_2-P_1$) finite
element pair. The turbine density function and the associated bottom
friction are represented by continuous, piecewise linear functions.

When solving a steady state problem, the time derivative terms are ignored in equations
\eqref{eq:to_spatially_discretised_shallow_water_eq}.
In the transient case,
equations~\eqref{eq:to_spatially_discretised_shallow_water_eq} are discretised
in time using the backward Euler scheme. Denoting time steps as superscripts, the
discrete shallow water equations are:
find $(\uvec, \eta) \in V_h\times Q_h$ such that for all $({\vec\Psi},
\Phi) \in V_{h}\times Q_{h}$:
\begin{equation}
\begin{split}
    \left\langle \frac{\uvec^n - \uvec^{n-1}}{\Delta t},  {\vec\Psi} \right\rangle_\Omega +
  \left\langle  \uvec^n \cdot \nabla  \uvec^n,  {\vec\Psi} \right\rangle_\Omega + \nu
  \left\langle\nabla  \uvec^n, \nabla  {\vec\Psi}\right\rangle_\Omega & \\
 + g \left\langle \nabla \eta^n,  {\vec\Psi} \right\rangle_\Omega  + \left\langle
 \frac{c_b + c_t(d)}{H^n} \| \uvec^n\|  \uvec^n,  {\vec\Psi}
 \right\rangle_\Omega &=  ~0, \\
 \left\langle \frac{\eta^n - \eta^{n-1}}{\Delta t}, \Phi \right\rangle_\Omega -
\left\langle H^n \uvec^n, \nabla \Phi \right\rangle_\Omega + \left\langle H^n
\uvec^n
\cdot  \nvec, \Phi\right\rangle_{\partial \Omega \setminus \partial \Omega_N} &=  ~0.
\end{split}\label{eq:to_time_and_spatially_discretised_shallow_water_eq}
\end{equation}

\subsection{Optimisation method}\label{sec:optimisation_method}
The non-discretised problem~\eqref{eq:smeared_optimisation_problem} is an
infinite-dimensional optimisation problem in the sense that the
turbine density function, the control variable, lives in an infinite-dimensional
space. The
dimension of the corresponding discrete optimisation problem,
based on the discretised system \eqref{eq:to_time_and_spatially_discretised_shallow_water_eq}, depends on the mesh and the finite
element choice. For practically relevant setups, this results in a large-scale optimisation problem;
for instance, the multi-farm design optimisation in
section~\ref{sec:pentland-firth-optimisation}
has $161,763$ control dimensions.
Considering that each profit evaluation requires solving the shallow water equations, the solution of
such optimisation problems is only feasible with optimisation methods
whose iteration number requirements do not depend on the dimension of the control space
and which compute the solution in a small (e.g. in the order of 100 or less) number of
iterations.

Optimisation methods can be classified into derivative-free and derivative-based
algorithms. Derivative-free methods, such as genetic algorithms,
simulated-annealing or Powell's conjugate direction method, rely purely on
evaluating the goal quantity at different points in control space. Derivative-free
methods are easily applicable because the hydrodynamic solver can be treated as a
black-box. However, they typically require a large number of iterations
and scale unfavourably with the dimension of the control space.
Hence these methods become quickly infeasible for the problem considered here.

In contrast, derivative-based optimisation methods use both evaluation of the goal quantity
and its derivative with respect to the control variables. As such,
a solver for hydrodynamic equations alone is not sufficient, because the efficient
computation of derivatives requires a solver for the corresponding adjoint equations (see section \ref{sec:adjoint_equations}).
Derivative-based methods are very efficient for large-scale problems, and often
require only a small number of iterations, independent of the problem discretisation.
One potential drawback of derivative based methods is that they only guarantee local convergence,
that is they might not be able to find the global optimal solution.
However, in our experience this is mostly a problem if the feasible set of control functions
is non-convex, which is avoided with the proposed problem formulation here (see last remark in section~\eqref{sec:form_opt_problem}).
In this paper we use the limited-memory BFGS method with bound constraints
(L-BFGS-B) to solve the optimisation problems \citep{byrd1995, zhu1997b}.

\subsection{The adjoint equations}\label{sec:adjoint_equations}
The L-BFGS-B method requires the gradient of the
profit function with respect to the turbine density function.
However, the profit function, \eqref{eq:profit} or \eqref{eq:smeared_optimisation_problem},
depends on the turbine density both directly as well as indirectly
through the dependence of the velocity on the turbine density.  One could
approximate the gradient through a finite difference like approach, but this would require as
many profit evaluations as there are unknowns in the turbine friction representation. Each evaluation requires solving the shallow water equations to provide $u$ for a given $d$, making this only feasible for small
problems.

In contrast, the adjoint approach computes the gradient by solving
only a single linear partial differential equation system, termed the adjoint equations corresponding to the underlying shallow water equations. This means that the computational
cost becomes practically independent of the problem size.
The gradient is computed in the following three steps:

\begin{enumerate}
    \item Solve the shallow water equations for $\uvec, \eta$
        (section~\ref{sec:discretisation}). Note that this computation needs to be performed regardless in order to evaluate the farm profit.
\item Solve the
adjoint shallow water equations for the adjoint velocity $\vec {\lambda}_{u}$ and
adjoint free-surface displacement $\lambda_\eta$:
\begin{equation*}
\begin{aligned}
- \frac{\partial  \vec {\lambda}_u}{\partial t} + (\nabla  \uvec)^*  \vec {\lambda}_u -
\left(\nabla \cdot  \uvec \right)  \vec {\lambda}_u -  \uvec \cdot \nabla  \vec {\lambda}_u - \nu \nabla^2  \vec {\lambda}_u \\
- H \nabla \lambda_\eta + \frac{c_b + c_t(d)}{H} \left(\| \uvec\|  \vec {\lambda}_u +
\frac{ \uvec \cdot  \vec {\lambda}_u}{\| \uvec\|}  \uvec\right) & = \frac{\partial
\textrm{Profit}}{\partial  \uvec}^*, \\
- \frac{\partial \lambda_\eta}{\partial t} - g \nabla \cdot
\vec {\lambda}_u -\nabla\lambda_\eta \cdot \uvec - \frac{c_b+c_t(d)}{H^2}
\|\uvec\| \uvec& = 0. \\
\end{aligned}\label{eq:turbine_optimisation_continuous_adjoint_equations}%
\end{equation*}
For brevity we
refer to \citet[appendix C]{funke2013} for the derivation of these equations.
Note that the
adjoint equations are linear even though the shallow water equations are
non-linear, and that they are
solved backward in time with a zero final-time condition. The Dirichlet boundary conditions are the
homogenised boundary conditions of the shallow water equations.

In order to obtain a discretely
consistent gradient (such that the gradient is the exact derivative of the discrete
model), the adjoint equations must use the same discretisation scheme as the
shallow water equations.
The discrete adjoint shallow water equations are thus:
find $(\vec {\lambda}_u, \lambda_\eta) \in V_{h}\times Q_{h}$ such that
for all $({\vec\Psi},
\Phi) \in V_{h}\times Q_{h}$:
\begin{equation*}
\begin{aligned}
    \left< \frac{\vec {\lambda}_u^n -
    \vec {\lambda}_u^{n+1}}{\Delta t}, {\vec\Psi} \right>_\Omega
    + \left<\vec {\lambda}_u^n, {\vec\Psi}\cdot \nabla \uvec^n\right>_\Omega
    +  \left<\vec {\lambda}_u^n, \uvec^n \cdot \nabla
    {\vec\Psi}\right>_\Omega \\
    + \nu \left<\nabla  \vec {\lambda}_u^n, \nabla {\vec\Psi}\right>
    - \left<H^n \nabla \lambda_\eta^n, {\vec\Psi}\right>_\Omega
    + \left\langle H^n \lambda_\eta^n,  {\vec\Psi} \cdot  \nvec
    \right\rangle_{\partial \Omega\setminus\partial\Omega_N} \\
    + \left<\frac{c_b + c_t(d)}{H^n} \left(\|
        \uvec^n\| {\vec\Psi} + \frac{ \uvec^n \cdot {\vec\Psi}}{\| \uvec^n\|} \uvec^n
    \right), \vec {\lambda}_u^n \right>_\Omega&
        = \left<\frac{\partial \textrm{Profit}}{\partial \uvec^n}, {\vec\Psi}\right>_\Omega, \\
    \left<\frac{\lambda_\eta^n - \lambda_\eta^{n+1}}{\Delta t}, \Phi\right>_\Omega
    + g \left<\vec {\lambda}_u^n, \nabla \Phi\right>_\Omega
    - \left<\nabla \lambda_\eta^n, \Phi \uvec^n\right>_\Omega \\
    + \left< \lambda_\eta^n, \Phi \uvec^n \cdot \nvec\right>_{\partial
    \Omega\setminus\partial\Omega_N}
    - \left<\frac{c_b + c_t(d)}{\left(H^n\right)^2} \Phi \|\uvec^n\| \uvec^n,
    \vec {\lambda}_u^n\right>
    & = 0. \\
\end{aligned}\label{eq:turbine_optimisation_discrete_adjoint_equations}
\end{equation*}
Note that the adjoint equations depend on the velocity solution at all timesteps.
Here, we apply an ``all-in-memory'' approach, where we store
all velocity solutions from step 1 in memory. For large-scale problems where
this becomes a bottleneck, a disk-based check-pointing scheme could be
used.
\item Evaluate the gradient of the goal quantity with
\begin{equation*}
    \frac{\textrm{d} \textrm{Profit}}{\textrm{d} d} = -\sum_n\left(\frac{C_T
    A_T}{2H^n}
    \|\uvec^n\|\uvec^n \cdot \vec {\lambda}^n_u \right) + \frac{\partial
    \textrm{Profit}}{\partial d}.
\end{equation*}
\end{enumerate}

\section{Continuous versus discrete farm optimisation}\label{sec:contvsdisc}
\citet{funke2014} proposed a \textit{discrete} farm layout optimisation approach
with the objective to find the individual turbine locations that maximise the farm's power production (so-called turbine micro-siting).
The drag force exerted by a turbine was represented via an
increased bottom drag in the shallow water model, but in
contrast to the continuous approach proposed here the turbines were
explicitly resolved in the numerical model/mesh. More precisely, the farm induced
drag consisted of a sum of individual friction `bump' functions, each representing one tidal
turbine.  In that case the control parameters for the optimisation were the locations
of the bump centres, that is the locations of the individual turbines.

Given the similar objectives of this work, it is natural to ask how the discrete
and continuous approaches compare. In particular, our goal is
to address the following questions: Do the discrete and continuous approaches yield
similar optimal farm configurations and profit predictions? Are the predicted
farm performances comparable? Is the optimal number of turbines computed by
the continuous optimisation consistent with the discrete approach?

In an attempt to address these we applied both strategies to the same idealised problem.
The domain considered is a square basin with sides of 4 km, and $50$ m depth.
The farm area is restricted to a
square with sides of 1 km located in the centre the domain.  For
the discrete turbine optimisation, the turbine positions are restricted to this
same area.  The boundary conditions are an inlet flow of $2\, \textnormal{m}\,\textnormal{s}^{-1}$ on the west
boundary, a free surface elevation $\eta=0$ on the east boundary and free-slip conditions on
the north and south boundaries. The remaining parameters are listed in
table~\ref{tab:smeared-vs-discrete-parameters}. With this setup, the flow in the
channel is steady, which means that a steady-state shallow water solver could be
used.

\begin{table}
  \begin{tabular}{llrl}
      \hline
      Parameter & Symbol & \multicolumn{2}{c}{Value} \\
      \hline
      Water density & $\rho$        & 1000 & $\textnormal{kg}\,\textnormal{m}^{-3}$ \\
      Viscosity & $\nu$ & 0.5 & $\textnormal{m}^2\,\textnormal{s}^{-1}$ \\
      Water depth at rest & $h$ & 50 & m \\
      Gravity & $g$ & 9.81 & $\textnormal{m}\,\textnormal{s}^{-2}$ \\
      Natural bottom friction & $c_b$ & 0.0025 \\
      Turbine diameter        &    & 20 & m \\
      Minimum distance between turbines & $D_\text{min}$ & 40 & m \\
      Maximum turbine density & $\bar{d}$ & $6.25 \cdot 10^{-4}$ & m$^{-2}$ \\
      Thrust coefficient & $C_T$    & 0.6 \\
      Turbine cross section & $A_T$ & 314.15 & m$^2$ \\
      Profit margin & $m$           & 40 & $\%$\\
      Cost coefficient (break even power) & $C/(LIk)$ & 452.39 & kW \\
      \hline
  \end{tabular}
  \caption{Parameters for the continuous versus discrete
  farm optimisation comparison.} \label{tab:smeared-vs-discrete-parameters}
\end{table}

\subsection{Continuous turbine optimisation}
We solved the continuous farm design problem with the steady-state shallow water equations.
The steady-state solution allows us to
compute $E/L$ directly as the power extracted according to
\eqref{eq:power_extracted}.  The cost coefficient is calculated using equation
\eqref{eq:cost-coefficient} without the $0.42$ factor and a value of
$u_{\textrm{peak}}$ that is equal to the inlet velocity, see table \ref{tab:smeared-vs-discrete-parameters}.  The
computational mesh consisted of $40,590$ elements with a typical element size
of 10 m in the farm area and 100 m elsewhere.

The problem was discretised and solved as discussed in section~\ref{sec:numericalsolution}.
The non-linear shallow water problem was solved
with Newton's method and a direct linear solver for the linear sub-problems and the
adjoint problem.  After 274 optimisation iterations the relative increase in
the goal quantity per iteration dropped below $2.2\times 10^{-6}$ at which point
the optimisation algorithm terminated.

The resulting optimal turbine density function is shown in figure
\ref{fig:contvssmeared_optimal_density_continuum}. The optimal number of
turbines is determined with equation~\eqref{eq:opt_number_turbine} and yields $N=152$.
The power generated is $89.2$ MW, or an average power production of 586 kW
per turbine.
The cost of the farm, measured as extracted power needed to break even for the
optimal number of turbines, is $68.8$ MW. The difference, $20.4$ MW is the
profit measured in units of power.

\subsection{Conversion of the turbine density function into
  turbines positions}
\begin{figure}
  \centering
  \begin{subfigure}[b]{0.32\textwidth}
      \centering
      \includegraphics[width=\textwidth]{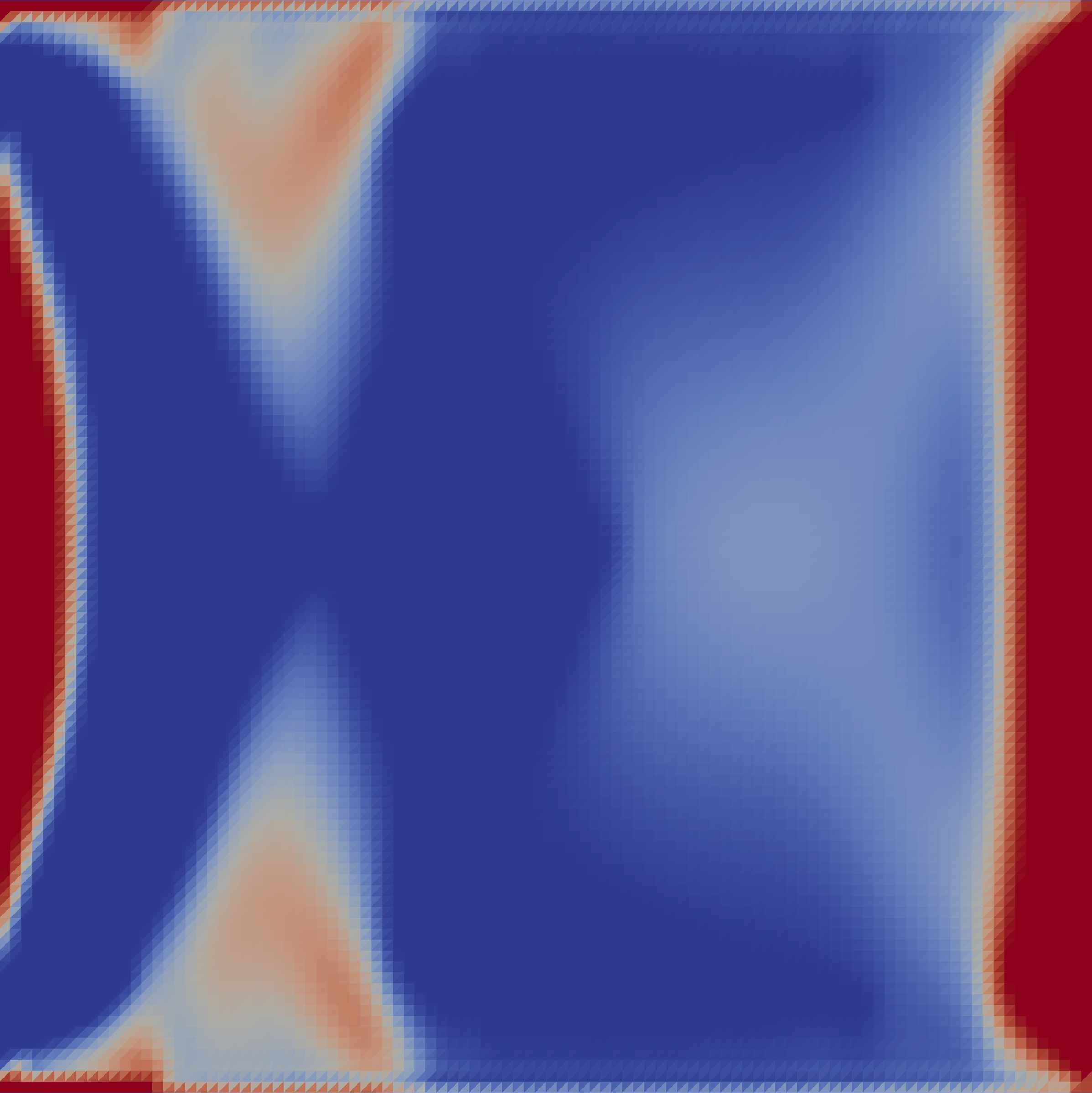}
      \caption{Continuous turbine approach.}
      \label{fig:contvssmeared_optimal_density_continuum}
  \end{subfigure}
  \begin{subfigure}[b]{0.32\textwidth}
      \centering
      \includegraphics[width=\textwidth]{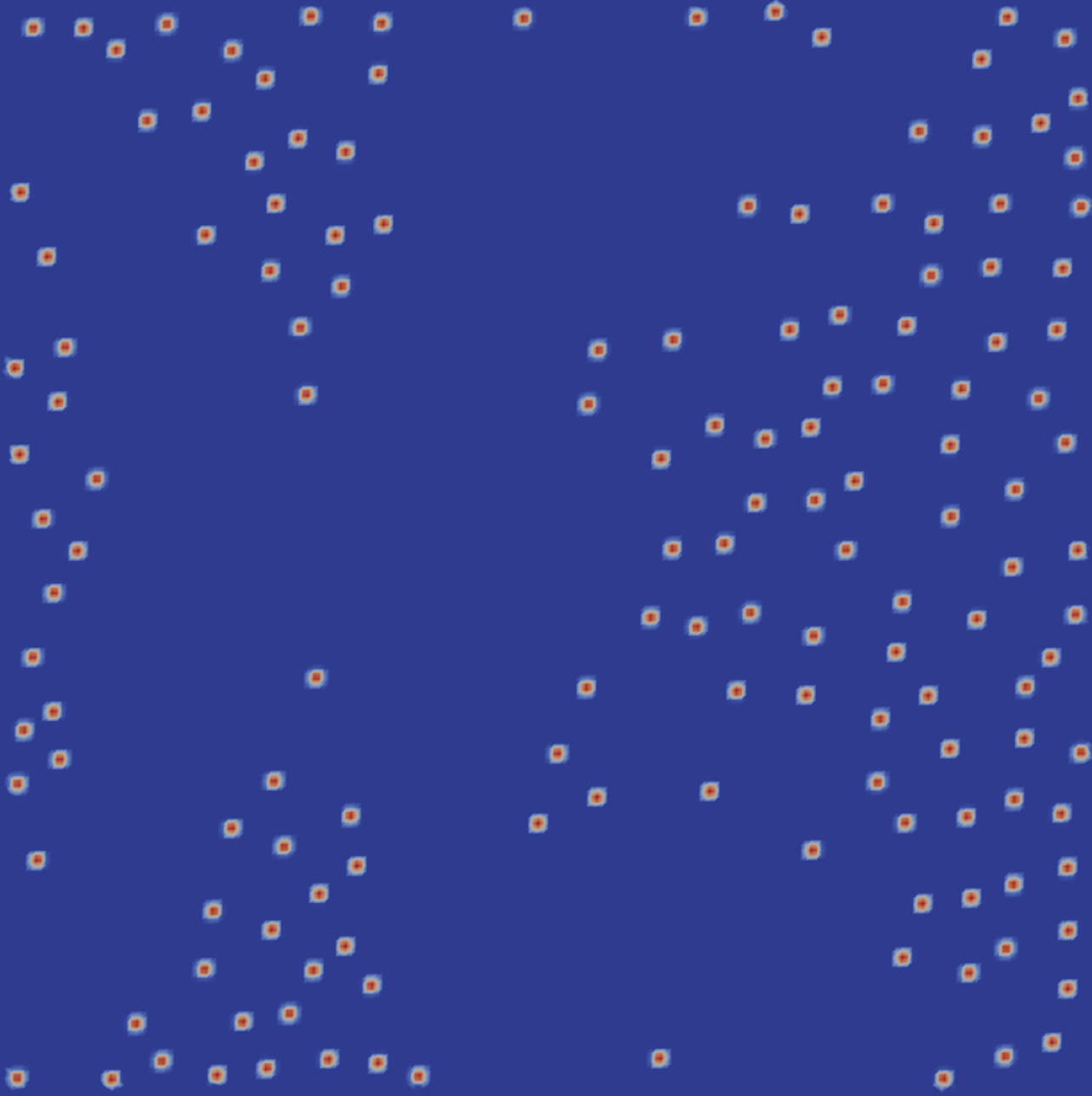}
      \caption{Discrete layout from continuous approach.}
      \label{fig:contvssmeared_optimal_density_discrete}
  \end{subfigure}
  \begin{subfigure}[b]{0.32\textwidth}
      \centering
      \includegraphics[width=\textwidth]{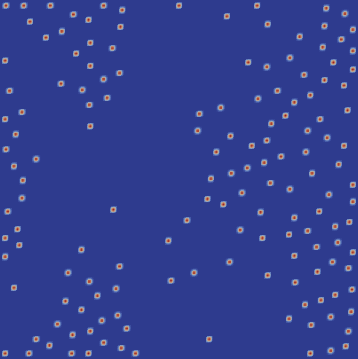}
      \caption{Discrete turbine approach.}
      \label{fig:contvssmeared_optimal_density_discrete_optimised}
  \end{subfigure}
  \caption{(a) Optimal turbine density field for the continuous versus discrete farm
      optimisation comparison. Only the farm
  area is shown. Blue indicates a farm density of $0$. Red indicates a farm
  density of the upper bound $\bar d$. The integrated turbine density field corresponds to $152$ turbines. (b)
  Positions of the individual turbines derived  from the optimal turbine density
  function using algorithm \ref{alg:conversion} with $N=152$. (c) Turbine
  positions after applying the discrete optimisation approach using (b) as
  an initial layout.}
  \label{fig:contvssmeared_optimal_density}
\end{figure}

Given the turbine density field obtained by the optimisation process,
we can derive individual positions for the
turbines as they are represented in the discrete approach via the following algorithm:

\begin{algorithm}{Conversion of turbine density function to turbine locations}\label{alg:conversion}
    \leavevmode
    \begin{enumerate}
        \item Determine the number of turbines, $N$, using equation
            \eqref{eq:opt_number_turbine}.
        \item Repeat until $N$ turbines have been deployed:
        \begin{enumerate}
            \item Pick a random point $\xvec$ in the domain.
        \item Place a turbine at $\xvec$ with probability $d(\xvec)/\max_\Omega \bar d$ if
            the minimum distance to all other turbines is at least $D_\text{min}$.
        \end{enumerate}
    \end{enumerate}
\end{algorithm}
Note that this algorithm only deploys turbines where $\bar d > 0$ and always generates layouts that satisfy the minimum
distance constraints between the turbines.
The result of the algorithm applied to the test problem is shown in
figure~\ref{fig:contvssmeared_optimal_density_discrete}. Comparing
figures~\ref{fig:contvssmeared_optimal_density_continuum} and
\ref{fig:contvssmeared_optimal_density_discrete} indicates that the algorithm
is able to capture most of the spatial structure from the turbine density function.

Next, we investigated how the continuous and discrete turbine representations
impact the hydrodynamic flow, and how much consistency there is in the predicted power and profit.
We configured the discrete turbine approach with the same settings as the
continuous turbine model. The turbine induced friction was set to a sum of bump
functions centred at the positions obtained from algorithm \ref{alg:conversion}. The
bump diameter was set to the turbine diameter and the bump amplitude was chosen such the integrated friction of the discrete
and continuous farm are identical. Since the discrete approach has to resolve
the bump functions, the mesh needed to be refined to $5$ m resolution over the farm
area.

\subsection{Turbine density function as an initial guess for discrete micro-siting}
The turbine locations obtained with algorithm \ref{alg:conversion} can additionally
be used to provide an initial guess for a discrete farm optimisation.
If the continuous approach generates good turbine concentrations, and consequently potential locations for individual turbines,
one perhaps would expect only marginal improvements from a subsequent optimisation
with the discrete approach. A consequence of this could also be that fewer iterations
of the more costly discrete optimisation may be needed. Hence, a combined
`continuous $\rightarrow$ discrete' process may provide broadly the same,
or potentially better, answer with a smaller overall computational cost compared
to using the discrete approach in isolation.

To test this, we performed two numerical experiments based on the idealised
problem considered earlier in this section.
Based on the value of $N$ obtained during the continuous optimisation above, the first experiment optimised the layout of $N=152$ turbines with the
discrete approach, starting from a regular non-staggered grid layout. In this case, the
initial power production was $41.4$ MW, which the discrete optimisation increased to
$84.6$ MW (the cost term remains constant as the number of turbines is
constant).

The second experiment used the turbine positions derived from the continuous
approach as an initial guess (as shown in figure
\ref{fig:contvssmeared_optimal_density_discrete}). In this case, the initial power extraction
was $79.4$ MW, i.e. just using the very simple algorithm \ref{alg:conversion} provides
a substantial improvement over a grid based initial layout, with a value within
approximately 12\% of that predicted by the continuous approach ($89.2$ MW). Following
discrete optimisation from this initial guess the optimised power extraction was
$84.5$ MW.  The optimisation
performs only small adjustments to the turbine positions and the final layout is
shown in figure~\ref{fig:contvssmeared_optimal_density_discrete_optimised}.

\begin{figure}[t]
  \centering
  \includegraphics[width=0.5\textwidth]{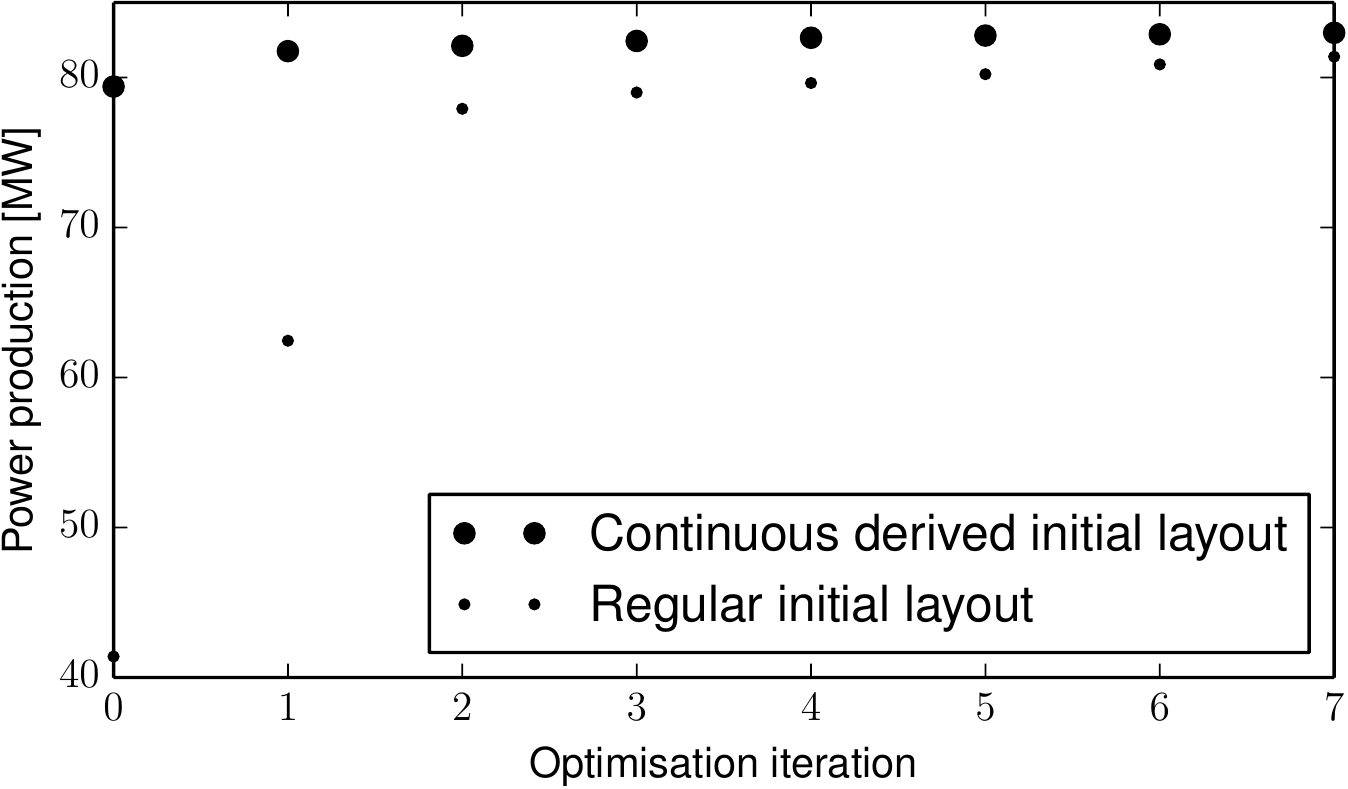}
  \caption{The power production during the first eight optimisation iterations for
  two different initial layouts.}
  \label{fig:opt_iter_for_density_as_initial_guess}
\end{figure}

Figure~\ref{fig:opt_iter_for_density_as_initial_guess} shows
an iteration plot of the first eight optimisation iterations for both experiments:
the initial guess derived from the continuous solution starts already close to the
optimal power production, and reaches the optimum practically after one
iteration. The optimisation with the initial regular layout converges to
essentially the same optimal power production, but requires more iterations.

These results suggest that the continuous approach may be directly used to provide
a valuable farm design in its own right. And further, if computationally feasible,
the locations can be further refined with a few iterations of the more costly discrete
optimisation.

\subsection{Comparison of hydrodynamic solutions}
\begin{figure}
  \centering
  \begin{subfigure}[b]{0.3\textwidth}
      \centering
      \includegraphics[width=0.9\textwidth]{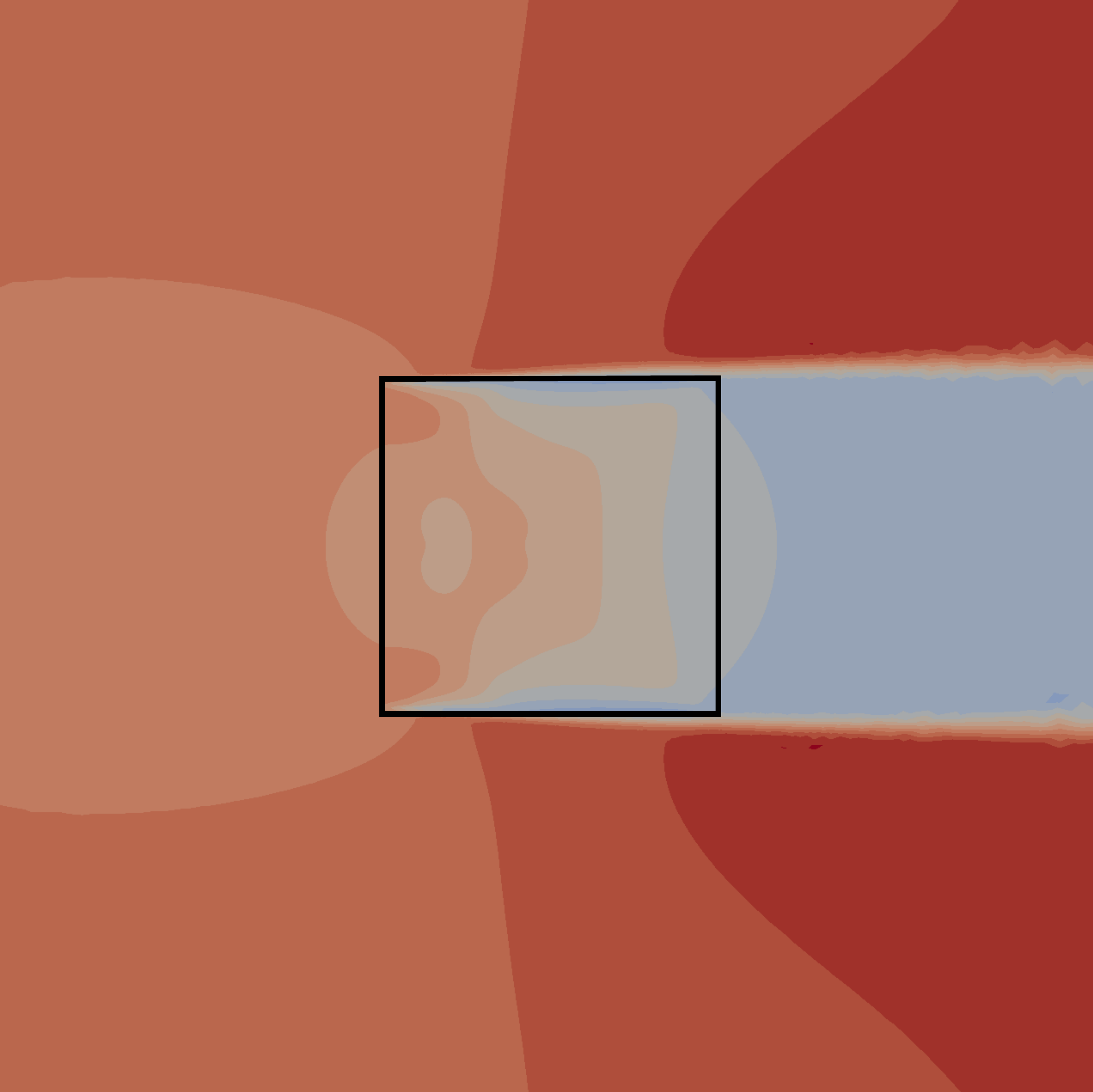}
      \caption{}
      \label{fig:svsd_flow_cont}
  \end{subfigure}%
  \begin{subfigure}[b]{0.3\textwidth}
      \centering
      \includegraphics[width=0.9\textwidth]{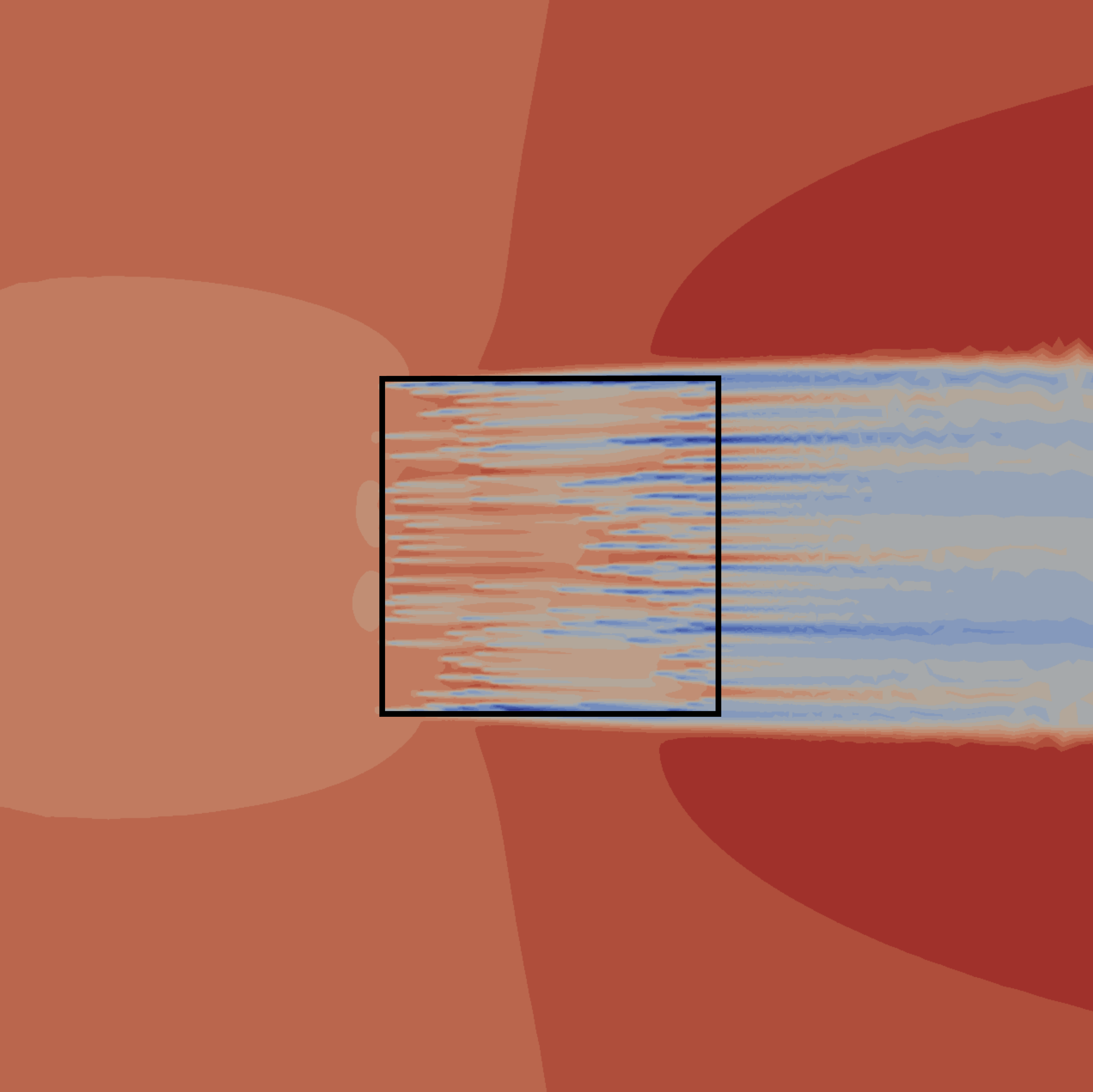}
      \caption{}
      \label{fig:svsd_flow_disc}
  \end{subfigure}%
  \begin{subfigure}[b]{0.3\textwidth}
      \includegraphics[width=0.9\textwidth]{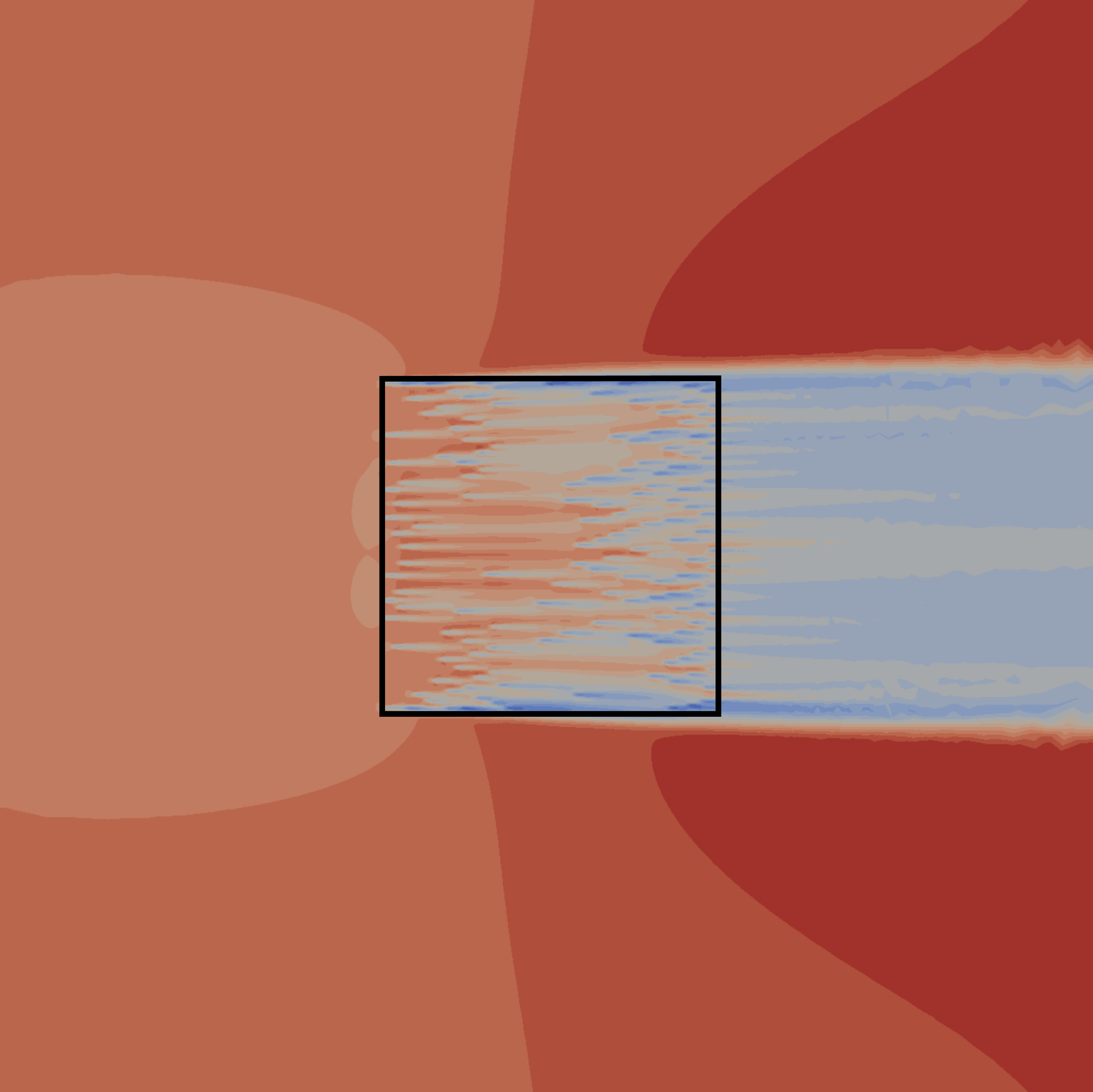}
      \caption{}
      \label{fig:svsd_flow_disc_opt}
  \end{subfigure}%
  \begin{minipage}[b]{0.1\textwidth}
    \vspace{3ex}
    \includegraphics[width=\textwidth]{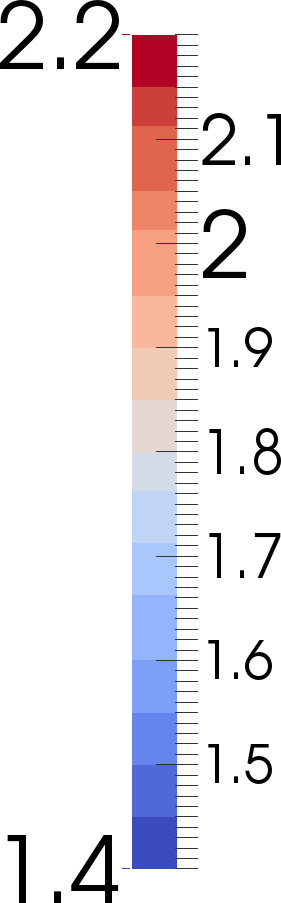}
  \end{minipage}
  \caption{Velocity magnitudes ($\textnormal{m}\,\textnormal{s}^{-1}$) in the presence of (a) the optimised continuous
    turbine farm, (b) the discrete turbine farm converted from the continuous
    optimisation result using algorithm 1, and (c) the discrete farm after
    further optimising using the discrete approach.}
  \label{fig:contvssmeared_optimal_density_vel}
\end{figure}

The hydrodynamic solutions for both turbine representations are compared in
figure~\ref{fig:contvssmeared_optimal_density_vel}. Figure
\ref{fig:svsd_flow_cont} shows the velocity magnitude in the presence of the optimised
continuous turbine farm, and figure \ref{fig:svsd_flow_disc} the solution with the
discrete farm derived directly from the continuous solution. Starting from this
layout as an initial guess, figure \ref{fig:svsd_flow_disc_opt} gives the flow solution after applying the discrete
approach to further optimise this farm.
As expected, the discrete approach shows
a much more detailed flow pattern inside the farm area, but with a broadly consistent
build up of a farm deficit within the farm area. In addition the
patterns of farm wake and bypass flow also match very well
with the pattern in the continuous approach. Going from figure
\ref{fig:svsd_flow_disc} to \ref{fig:svsd_flow_disc_opt}, the discrete solution gets even closer to
the continuous solution in figure \ref{fig:svsd_flow_cont} as the discrete farm is further
optimised.
This indicates that the continuous
approach gives a good approximation of the large scale flow pattern when
optimising turbine arrays.

\subsection{Optimal number of turbines}
As already stated, a significant strength of the continuous approach is that it
also predicts the optimal number of turbines for a farm
site. We performed a series of experiments to check if this number is also considered
to be optimal in the discrete approach.

First, we determined the optimal number of turbines $N =152$ with equation \eqref{eq:opt_number_turbine}.
Then, we applied algorithm \ref{alg:conversion} to generate initial layouts for the discrete
approach with $N, N \pm 25$ and $N \pm 50$ turbines, and performed a
discrete layout optimisation for each of them. The results are listed in table
\ref{tab:smeared-vs-discrete-results}.  The table shows that the discrete
optimisation with 152 turbines produces the highest profit. Therefore the
continuous approach indeed correctly predicted the optimal number of turbines
(to within a tolerance of around 16\% based on the numbers considered in this
particular case).

\begin{table}
    \centering
  \begin{tabular}{cc|cccc}
      \multicolumn{2}{c}{Experiment} & Power (MW) & Cost (MW)
    & Profit (MW) \\
    \hline
    \multicolumn{2}{c}{\textbf{continuous}}                    &  \textbf{89.21} &  \textbf{68.81} &
    \textbf{20.39} \\
\hline
discrete & N=102 & 59.89   &    46.18  & 13.71 \\
discrete & N=127 &  72.61  &    57.49 & 15.12 \\
\textbf{discrete} & \textbf{N=152} &  \textbf{84.45}  &    \textbf{68.81} & \textbf{15.64} \\
discrete & N=177 &  95.03  &    80.12 & 14.91 \\
discrete & N=202 & 104.20  &    91.44 & 12.76
  \end{tabular}
  \caption{Results from the continuous versus discrete farm optimisation comparison. The
  continuous approach predicts that 152 turbines yields the maximum profit. This is
  confirmed by running a set of discrete optimisation runs with varying number
  of turbines. The continuous approach results in a slightly higher predicted power extraction
  because its control space has more degrees of freedom than the discrete
  approach.}
  \label{tab:smeared-vs-discrete-results}
\end{table}

\section{Setup and validation of the Pentland Firth forward
model}\label{sec:pentland}
  \begin{figure}
    \includegraphics[width=\textwidth]{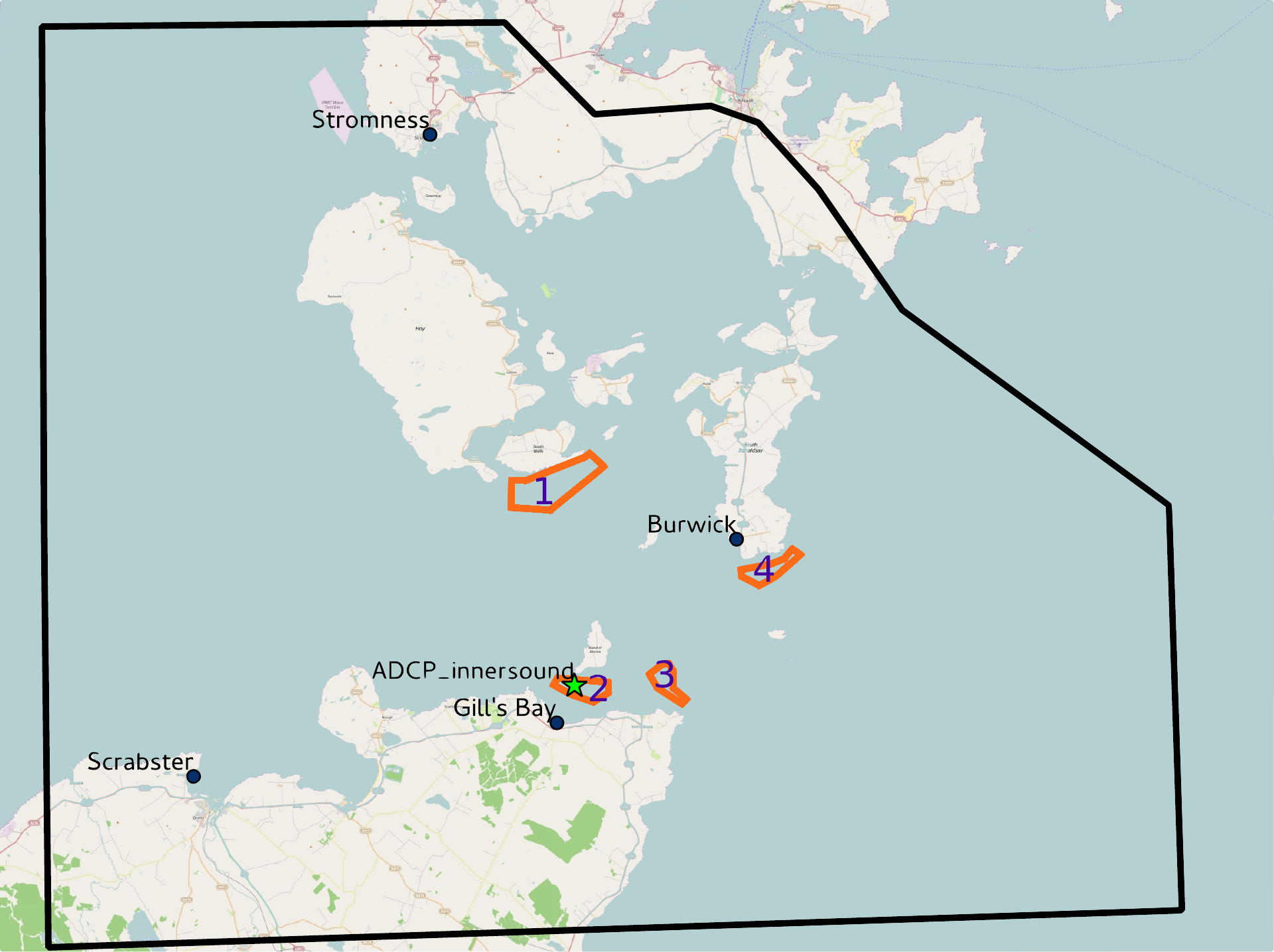}
    \caption{The Pentland Firth between the Scottish mainland and the Orkney
      Islands. The outline of the modelled area is indicated in black. For the model
      validation, data from four tide gauge locations at Scrabster, Stromness, Gill's Bay
      and Burwick, and from an ADCP in the Inner Sound have been used.
      The farm areas optimised in section \ref{sec:multi-farm-optimisation}
      are indicated and numbered: 1) Cantick Head, 2) Inner Sound, 3) Ness of
Duncansby, and 4) Brough Ness.}
    \label{fig:outline_and_gauges}
  \end{figure}

The examples in sections \ref{sec:pentland-firth-optimisation} and
\ref{sec:complex} to follow are based on a forward model of the
Pentland Firth, a strait between the Scottish mainland and the Orkney Islands.
It is an area of significant interest for the tidal renewable industry with several
leasing areas assigned by the Crown Estate. In one of these, in the Inner Sound
to the south of the island of Stroma, the construction and deployment of an initial
small number of turbines has been approved and should be underway soon.
This initial deployment is hoped to be scaled up to a large farm with a power production of 398 MW.

The model of the Pentland Firth covers a domain shown in figure
\ref{fig:outline_and_gauges}. Two different meshes were employed both with edge
lengths varying between 350 m and 25 m. The smallest edge lengths were only
employed inside the designated farm areas, with resolution
smoothly coarsening outside of these. The mesh used in section
\ref{sec:pentland-firth-optimisation} consists of
321,224 elements and
defines four farm areas which
are roughly based on four different lease sites assigned by
the Crown Estate \citep{crown2013}: 1) Cantick Head, 2) Inner Sound, 3) Ness of
Duncansby, and 4) Brough Ness. The mesh used in section \ref{sec:complex} only
incorporates the Inner Sound site (farm 2) as a high resolution area, and thus
consists of considerably fewer elements (115,022). Outside of the farm area(s) the
mesh resolutions are chosen in the same manner.
 The validation shown here is based on the mesh from section \ref{sec:complex}.
As expected, the results with the mesh from
section \ref{sec:pentland-firth-optimisation} were found to be very similar.

It should be noted that the chosen domain is smaller than the recommendations in
\citet{adcock2011}, which suggests the open boundaries should be a 100 km to the
east of Muckle Skerry at eastern exit of the Pentland Firth, and 200 km to the
west of the western entrance. Our choice was a compromise to limit the
computational expense. We did however check for differences in flow velocities
at the open boundaries between the base case without turbines and a case with
all four farms present (see
section \ref{sec:pentland-firth-optimisation}), and found the differences to be
small
(less than $0.02~\mathrm{ms}^{-1}$ at any point along the open boundaries for the entire
simulation time).

The bathymetry used in both setups is obtained from four different sources: 1)
the global bathymetry data set GEBCO\_08 \citep{gebco08_2010} with a resolution of 30 arc-sec
($\approx 900$ m), 2) Digimap \citep{digimap2014} which gives a resolution of 1
arc-seconds ($\approx 30$ m) near the UK
coast, 3) bathymetry data from the \citet{scottish_bathymetry2009} at 2 m resolution covering most of the Pentland
Firth, but missing parts of the Inner Sound near the coast, and 4) bathymetry data obtained from MeyGen, also at 2 m resolution but
covering a larger part of the Inner Sound. These four sources were combined to
obtain the most reliable data in each area. The result was then projected onto
the mesh and smoothed.

The models are forced with free surface elevation boundary conditions reconstructed from
tidal harmonics based on a global tidal solution using the OSU Tidal Prediction
Software (OTPS, \citet{otps2010}). At the coastal boundaries a no-slip condition
was applied. No wetting and drying scheme was applied here, but a minimum depth of
10 m constraint was applied to the bathymetry used in order to avoid negative water depths. The models use a fixed eddy viscosity of
10 $\textnormal{m}^2\,\textnormal{s}^{-1}$ everywhere, except for a band of 1 km around the open boundaries
where an increased viscosity of 100 $\textnormal{m}^2\,\textnormal{s}^{-1}$ is applied to avoid instabilities
typically arising at open boundaries in nonlinear models. For time integration
the backward Euler method was applied with a time step of 600~s.

For the validation of the models, the setup was run for a period of 20 days starting
on the 4th of July 2011 at midnight. This period was chosen to overlap with the
ADCP data available. The first day of model output was not used in the analysis
to allow for the
model to spin up.

\begin{table}
\begin{tabular}{l|rrrrrrrr}
   & \multicolumn{2}{c}{M2} & \multicolumn{2}{c}{S2} & \multicolumn{2}{c}{K1} &
  \multicolumn{2}{c}{O1}\\
   & amp & pha & amp & pha & amp & pha & amp & pha\\
  \hline
  {\bf Charted} \\
  Scrabster & 1.35 & 247 & 0.51 & 280 & 0.14 & 144 & 0.10 & 349 \\
  Stromness & 0.89 & 270 & 0.35 & 303 & 0.11 & 148 & 0.10 & 5 \\
  Gill's Bay & 1.12 & 268 & 0.41 & 300 & 0.13 & 151 & 0.11 & 360 \\
  Burwick & 0.88 & 287 & 0.35 & 322 & 0.14 & 158 & 0.12 & 17 \\
  \hline
  {\bf Model ($c_b=0.0025$)} \\
  Scrabster & 1.27 & 247 & 0.42 & 290 & 0.14 & 157 & 0.11 & 359 \\
  Stromness & 0.94 & 270 & 0.31 & 313 & 0.13 & 167 & 0.11 & 14 \\
  Gill's Bay & 0.96 & 276 & 0.31 & 318 & 0.13 & 166 & 0.11 & 11 \\
  Burwick & 0.86 & 291 & 0.27 & 335 & 0.13 & 167 & 0.11 & 15 \\
  \hline
  {\bf Model ($c_b=0.003$)} \\
  Scrabster & 1.27 & 247 & 0.42 & 290 & 0.14 & 157 & 0.11 & 359 \\
  Stromness & 0.95 & 270 & 0.31 & 312 & 0.13 & 167 & 0.10 & 14 \\
  Gill's Bay & 0.96 & 275 & 0.31 & 318 & 0.13 & 166 & 0.11 & 12 \\
  Burwick & 0.86 & 291 & 0.27 & 335 & 0.14 & 167 & 0.11 & 16 \\
  \hline
\end{tabular}
\caption{}
\label{tab:gauge-harmonic-analysis}
\end{table}

At four different locations in the domain (Scrabster,
Stromness, Gill's Bay and Burwick, see figure \ref{fig:outline_and_gauges}), the free
surface solution was recorded. From these time series the four main harmonic
constituents (M2, S2, K1 and O1) were derived using the T\_Tide
 Harmonic Analysis Toolbox \citep{pawlowicz2002} in MATLAB. The results are shown
 in table \ref{tab:gauge-harmonic-analysis} where they are compared with values
 based on measurements at these locations \citep{ukho2009}. The main driver of
 the tidal currents through the Pentland Firth is the phase difference across
 the channel. It is therefore important to get the phases correct in the
 different locations along the channel. From the table it becomes clear that
 there is good agreement for the main M2 constituent and reasonable agreement
 for S2. These results are very similar to those presented in \citet{Easton2012} and
 \citet{martin2015}.

\begin{figure}
  \includegraphics[width=0.5\textwidth]{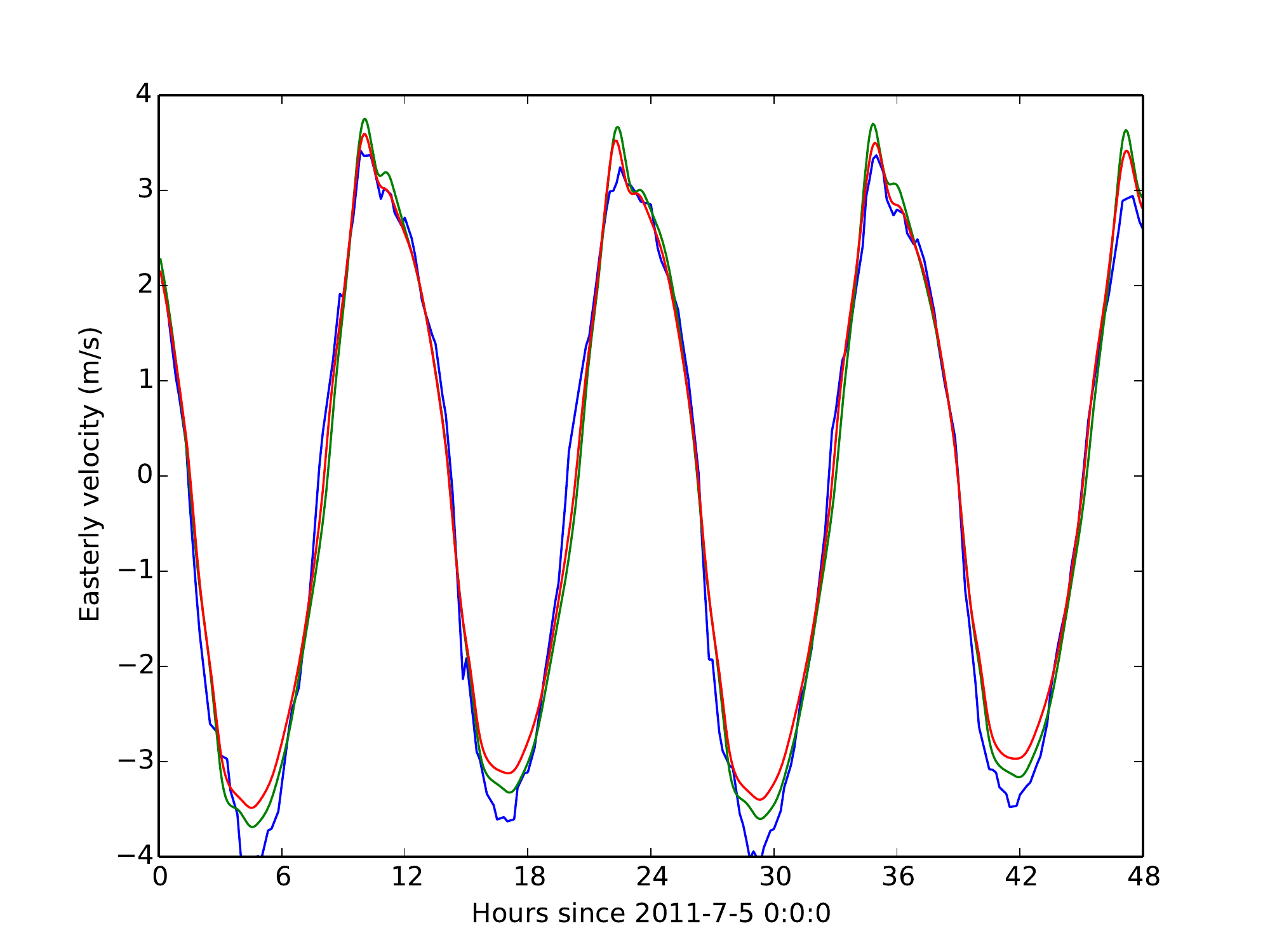}
  \includegraphics[width=0.5\textwidth]{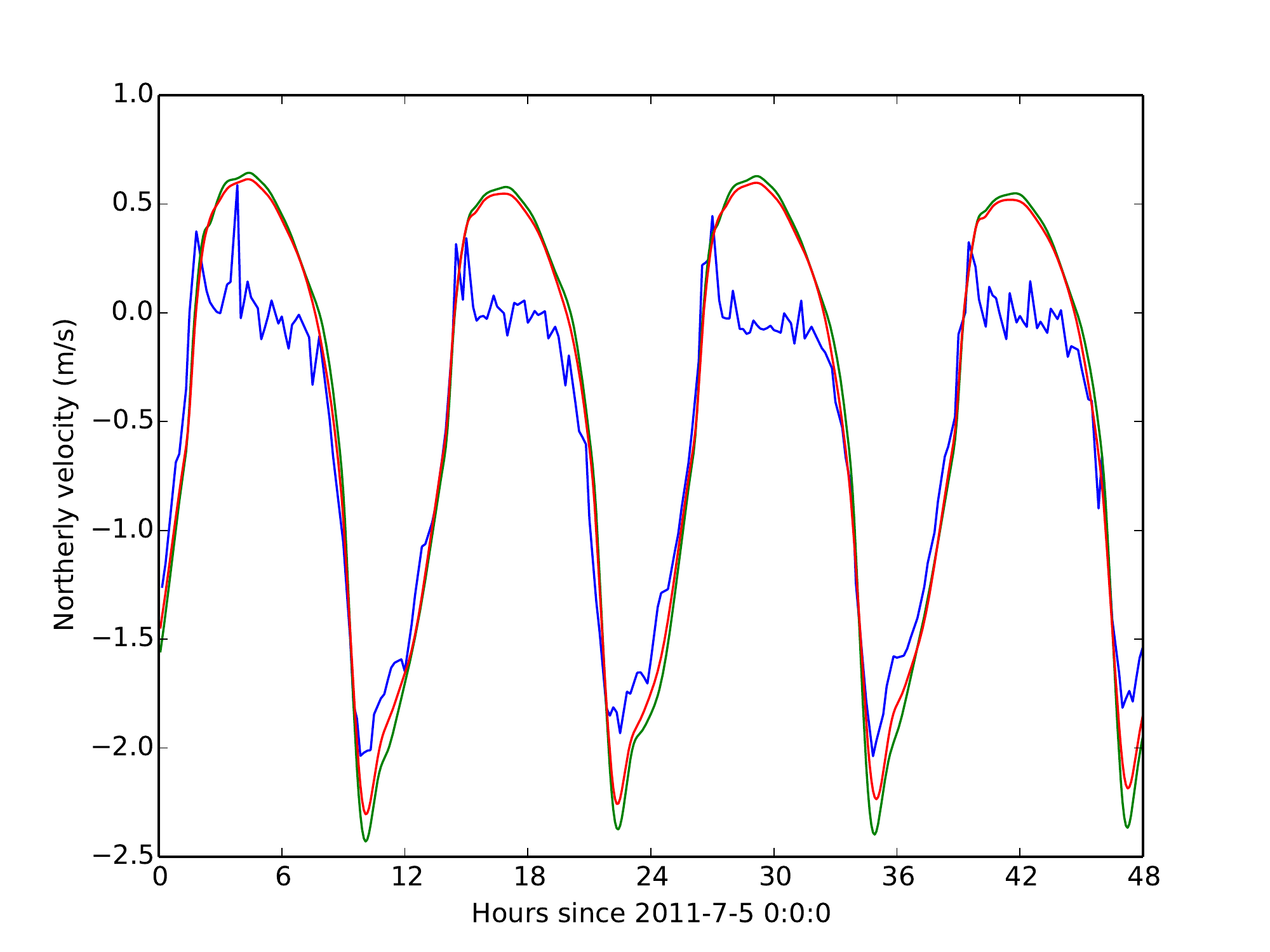}
  \caption{A comparison between both components of velocity (left figure: easterly,
    right figure: northerly) measured by an ADCP inside
    the Inner Sound farm (blue), and OpenTidalFarm model output for two values of bottom friction:
    $c_b=0.0025$ (green) and $c_b=0.003$ (red). The figure compares the depth
    average of the measured ADCP profile with the outcomes of the shallow water
    model.}
  \label{fig:adcp_comparison}
\end{figure}
For a validation of the predicted currents the measured ADCP
data from a location in the Inner Sound was available. It is located inside
farm area 2 which will be used in the optimisation runs in the following sections. In figure
\ref{fig:adcp_comparison}, we display both components of velocity for the first
two days of the analysis period, comparing the model
results with the ADCP measurements. The ADCP data was available at averaging
intervals of 10 min which will have filtered out the very high frequency signal.
It can however be observed that the signal still contains a lot of relatively
high frequency oscillations, that are particular strong during the ebb (westward
flow) period. These can be ascribed to the turbulent nature of the flow. The
main differences with the model results can be seen during the same ebb
period: where the model shows a consistent, small northerly component turning
the mainly westward current a few degrees northward, the ADCP shows a strongly
oscillatory signal with the current direction alternating around the
westward direction.

\newcommand\head[1]{\begin{minipage}{0.11\textwidth}
    \centering\scriptsize
    #1
\end{minipage}}
\begin{table}
  \setlength{\tabcolsep}{0.5ex}
  \begin{tabular}{l|ccccccc}
    &
    \head{Speed Bias Ebb (m/s)} &
    \head{Speed Bias Flood (m/s)} &
    \head{Speed RMS Error (m/s)} &
    \head{Speed Scatter Index} &
    \head{Speed Correlation Coefficient} &
    \head{Direction Bias Ebb (deg)} &
    \head{Direction Bias Flood (deg)} \\
    \hline
    $c_b=0.0025$ & -0.06 & 0.20 & 0.36 & 0.16 & 0.94 & -12 & -1 \\
    $c_b=0.003$ & -0.19 & 0.13 & 0.32 & 0.14 & 0.95 & -12 & -1
  \end{tabular}
  \caption{Statistics of the comparison, over 19 days, between ADCP and model outcomes for two
    different values of bottom friction: $c_b=0.0025$ and $c_b=0.003$. See
    \citet{martin2015} for definitions.}
  \label{tab:adcp_comparison}
\end{table}
In table \ref{tab:adcp_comparison} the main statistics of the comparison
between ADCP and model outcomes are given. Here 19 days of simulation time are
compared. The table confirms the bias of 12 degrees northward in the direction
of the current, and an underprediction in speed during the ebb period that was
also observed for the first two days in figure \ref{fig:adcp_comparison}. Of the two
values for the background bottom friction coefficient, $c_b=0.0025$ and
$c_b=0.003$, that were considered, slightly better results are produced with
$c_b=0.003$. This value was therefore chosen in the following sections. As expected the bottom friction coefficient has less of an impact
on the free surface elevations as becomes clear from table
\ref{tab:gauge-harmonic-analysis}.

\section{Pentland Firth farm design optimisation}\label{sec:pentland-firth-optimisation}
\subsection{Individual farm optimisation}\label{sec:individual-farm-optimisation}

We applied the continuous farm design approach separately to the four tidal farm sites in
the Pentland Firth, UK (figure \ref{fig:outline_and_gauges}).
The computational domain was equivalent to the domain discussed in section
\ref{sec:pentland}. The mesh element sizes ranged between 350 m and 200 m
outside the farm areas, and 25 m in the farm area. The simulation
time spans over one tidal cycle ($12.5$ h), starting at 5:01:35am on
5 July 2011, with a 5 min time step. The initial velocity condition was taken from the
simulation discussed in section \ref{sec:pentland}, to avoid the need for a
spinup period. The starting time was
chosen, between subsequent slack tides, such that the impact of changes in
friction in the farm area would have minimal impact on the current starting from
the initial condition.
 The remaining parameters are equivalent to the ones discussed in section
\ref{sec:pentland}.
Turbine characteristics were chosen to be typical for turbines expected to be
installed in the area: a turbine diameter of $20~\textrm{m}$, i.e.
$A_t=314~\textrm{m}^2$, was chosen with a
fixed thrust coefficient of $C_t=0.6$, which is significantly
lower than the optimal value of 8/9 but more realistic for large scale
turbines in operating conditions. For comparison a Hammerfest HS-1000 at rated
speed of $2.7~\textrm{ms}^{-1}$ has a published \citep{hammerfest12} rated power
of $1~\textrm{MW}$, whereas
our idealized turbine produces $1.5~\textrm{MW}$ at this speed. For simplicity we do not
include rating in our calculations, but assume a constant thrust coefficient
(see section \ref{blockage_and_rating} for a discussion).
This means that at higher speeds we extract an unrealistic amount of
power, e.g. $5~\textrm{MW}$ at $4~\textrm{ms}^{-1}$.

The cost coefficient, that is the average power production per turbine to
break even, was derived from \eqref{eq:cost-coefficient} with a thrust
coefficient $C_T=0.6$ and
turbine cross section $A_T=314~\text{m}^2$ as mentioned, and water density $\rho = 1000~\text{kg}\,\text{m}^{-3}$, profit margin $m=73~\%$,
and peak velocity $u_\text{peak} = 3.5~\text{m}\,\text{s}^{-1}$. This yields $C/LIk = 452$ kW.
Note that the cost coefficient is expected to have a large impact on the optimal
number of turbines. The minimum distance between two turbines is set to
$D_\text{min} = 40$ m,
which results in a maximum turbine density of $\bar d = 6.25 \times
10^{-4}~\textrm{m}^{-2}$ in the farm.

For each farm, we terminated the optimisation after $50$ optimisation
iterations. On average, this required $54$ shallow water  and $50$ adjoint
shallow water solves.  At this point there was no visible change in the farm
design and the relative change in the profit function per iteration was
always below $0.001 \%$. For testing purposes, more iterations were performed
but yielded indistinguishable results.

The optimal designs for the four farms are shown in
figure~\ref{fig:multi_farm_optimisation_results_layouts} (left images).
The following key characteristics in the optimal designs can be observed:
\begin{itemize}
    \item The optimal designs differ significantly between the farms, confirming
        that site-specific design optimisation is important.
    \item The turbines are not evenly spread across the farm, but rather
        concentrate around a few areas with high turbine densities.
    \item A common pattern in the optimal design are barrage-like structures
    aligned perpendicular to the flow.
    \item It can be beneficial to deploy turbines along the site boundary.
\end{itemize}

The striking differences between the different sites can be explained by
differences in the local flow characteristics, in particular the alignment
of the farm area with the streamlines of the flow. As an example, the optimal
solution for farm 2 in the Inner Sound of Stroma, seems to suggest a double row
of turbines in the eastern half of the farm. The location of the first (western)
of these two fences is probably guided by a principle to maximize the width over
which the turbines can block the flow including the most southern streamlines
that can be reached from within the farm area. The position of the second
(eastern) fence is constrained by the fact that putting it further east would
limit the width of the stream tube that is captured, and keeping enough distance
from the first one. The middle fence is placed in such a way that the streamline
furthest north is reached and the accelerated flow in the jet that comes off the
tip of the island is captured. Finally, a fence on the far west of the farm area
is suggested. These considerations demonstrate that for the design of a lease
area, which is typically guided by various other constraints such as
bathymetry and other commerical usage, a good understanding of the local tidal
flow is also essential.

The profit, power, cost, and number of turbine predictions are listed in
table~\ref{tab:multi_farm_results} (top four rows). Note that for consistency
with the next section, an additional hydrodynamic run was performed to diagnose
these values, with all four optimal farm designs incorporated in the same run, even
though their designs had been obtained without the presence of other farms.
The design of each farm reaches its
maximum profit with the deployment of hundreds of turbines.
In this experiment, farm 2 is arguably the most
promising, in terms of the number of turbines, the average power production, and profit.
In contrast farm 1 yields a relatively small profit, despite its large site area and
large number of turbines.

\subsection{Simultaneous multi-farm optimisation}\label{sec:multi-farm-optimisation}

The farm design optimisations conducted so far assumed that no other farm is
installed within the immediate vicinity. Since there exist only a limited number of tidal sites with strong currents,
multiple site developers often plan tidal farms in relatively close
proximity to each other. For the site developers it is therefore of practical
importance to understand how tidal farms influence each other, and the
impact of any interactions in terms of the optimal designs and profit predictions \citep{draper2013, draper2014a}.

To study the influence of multiple farms, we performed another design
optimisation, in which the design of all four farms were optimised
\textit{simultaneously} to maximise the overall profit. The setup is otherwise
identical to the previous section.

\begin{figure}[t]
  \centering
  \begin{subfigure}[b]{0.48\textwidth}
    \includegraphics[width=\textwidth]{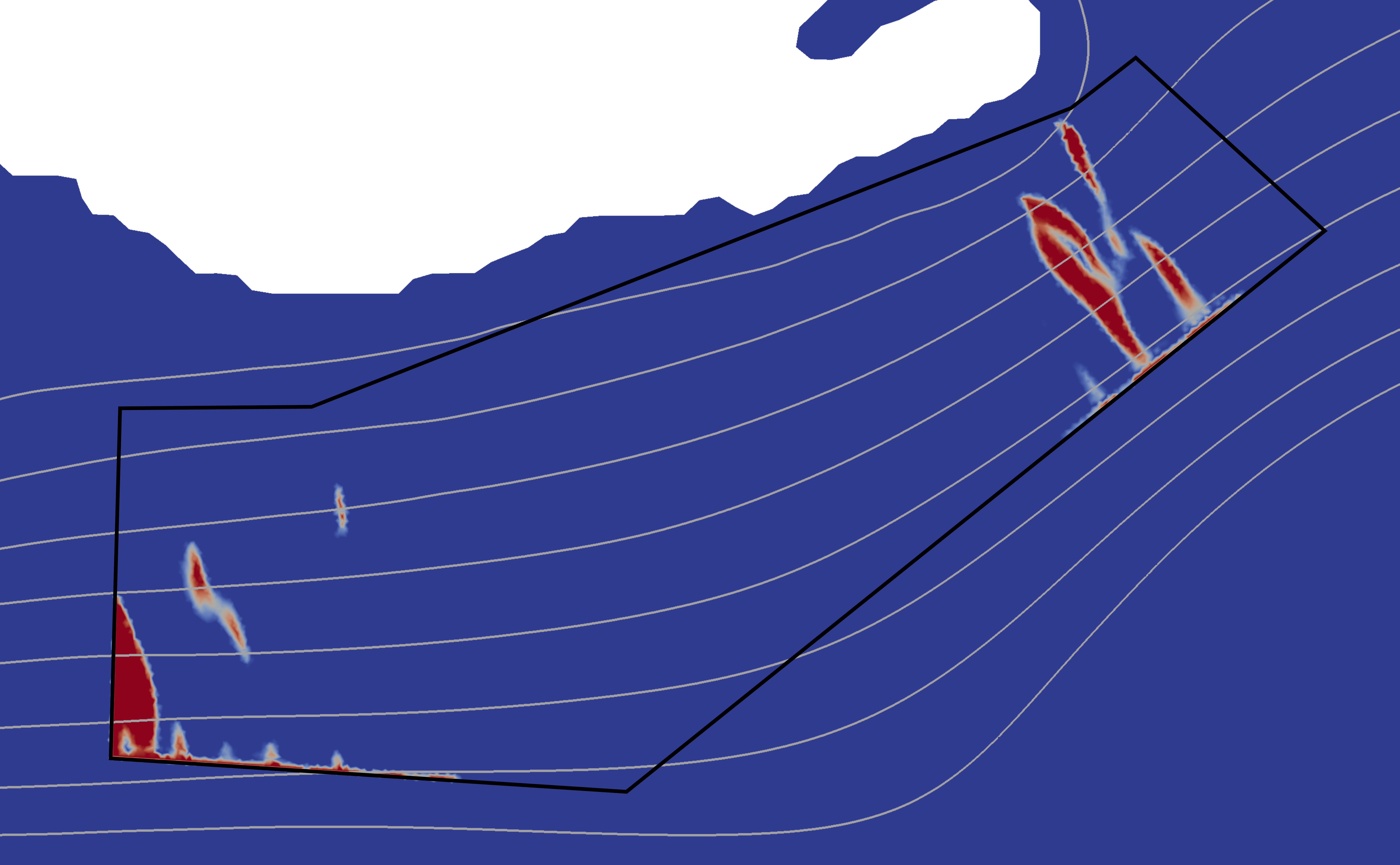}
    \caption{Farm 1 (228 turbines)}
    \label{fig:multi_farm_optimisation_farm_1_greedy}
  \end{subfigure}
  \begin{subfigure}[b]{0.48\textwidth}
    \includegraphics[width=\textwidth]{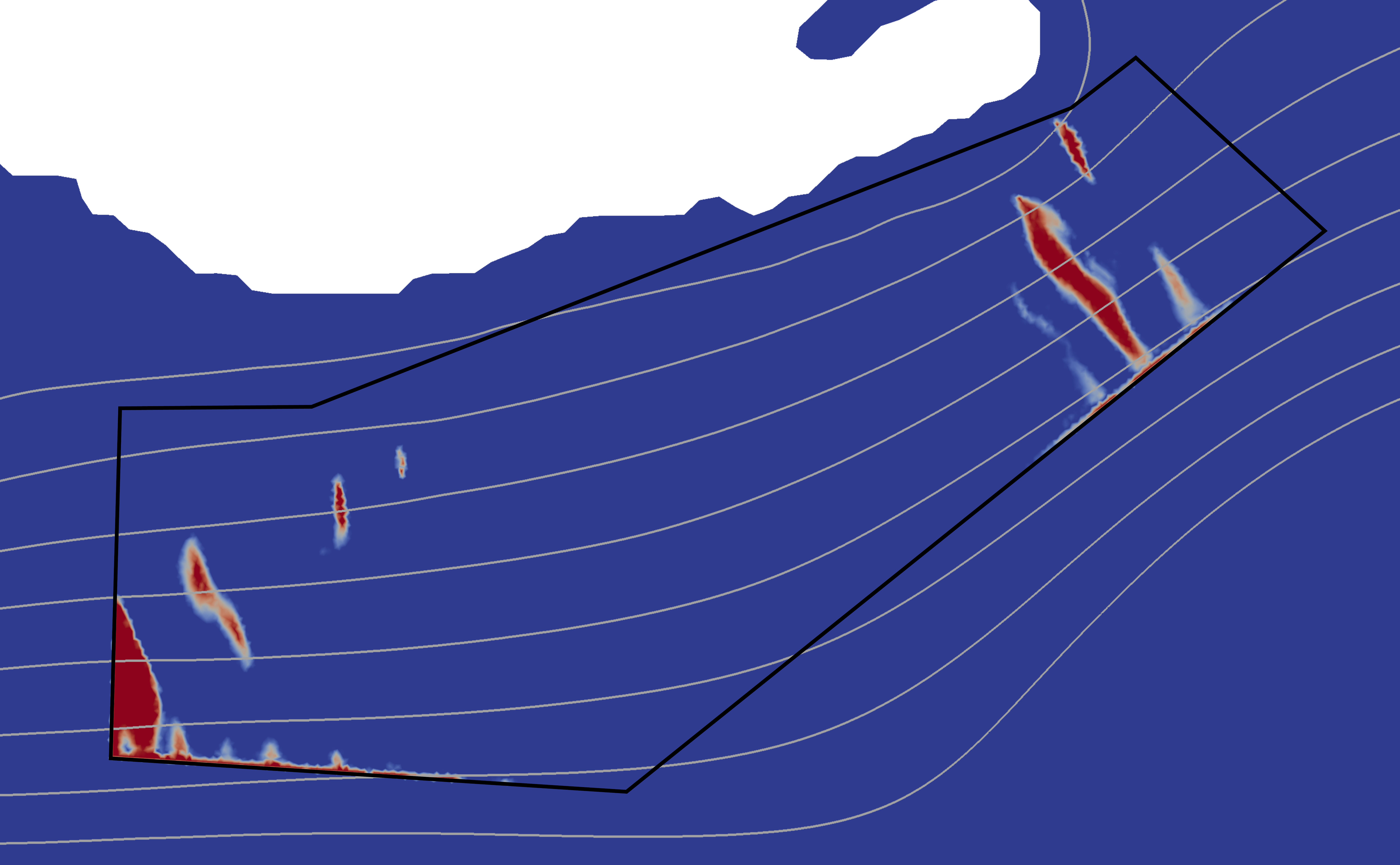}
    \caption{Farm 1 (224 turbines)}
    \label{fig:multi_farm_optimisation_farm_1_team}
  \end{subfigure}
  \begin{subfigure}[b]{0.48\textwidth}
    \includegraphics[width=\textwidth]{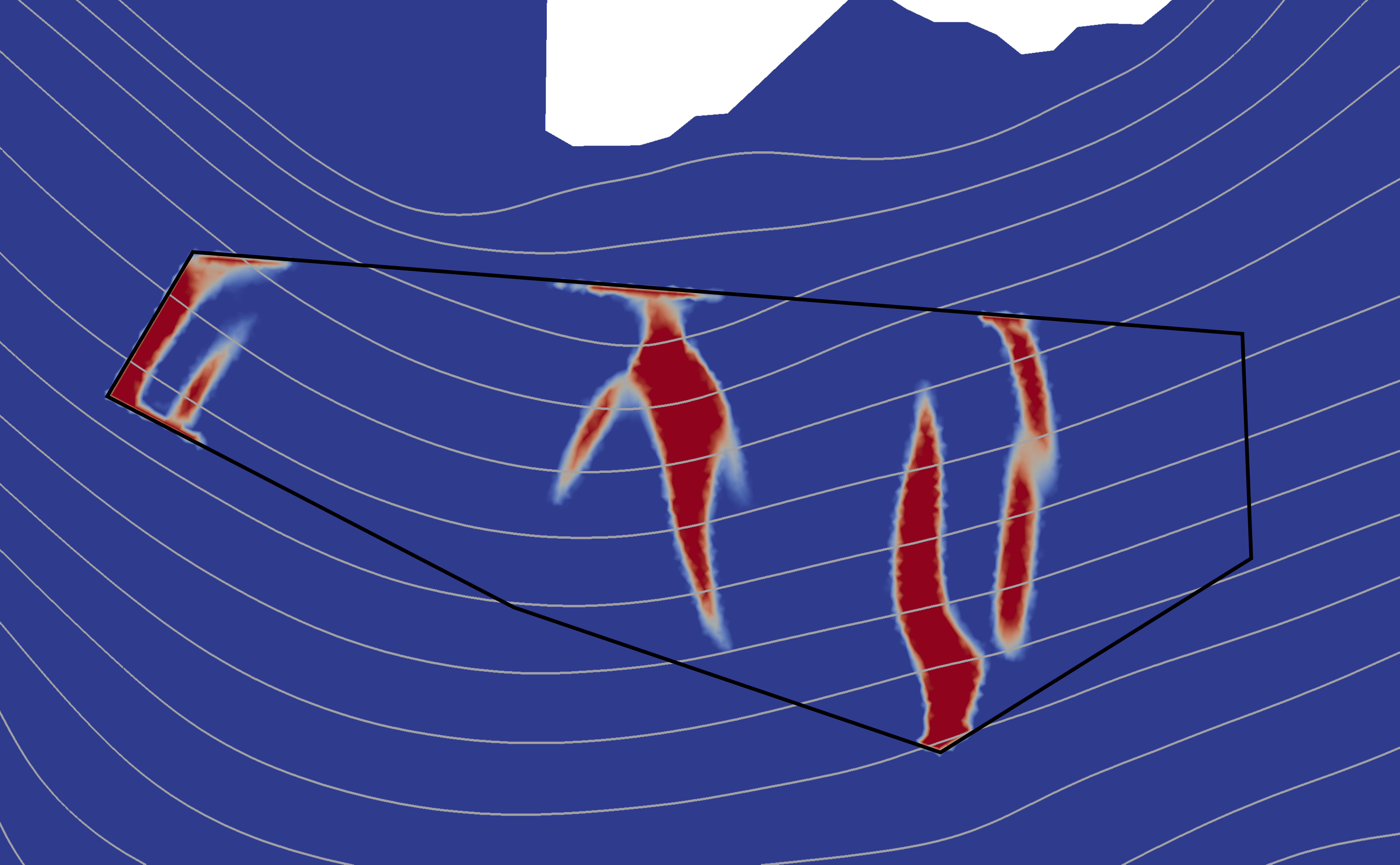}
    \caption{Farm 2 (266 turbines)}
    \label{fig:multi_farm_optimisation_farm_2_greedy}
  \end{subfigure}
  \begin{subfigure}[b]{0.48\textwidth}
    \includegraphics[width=\textwidth]{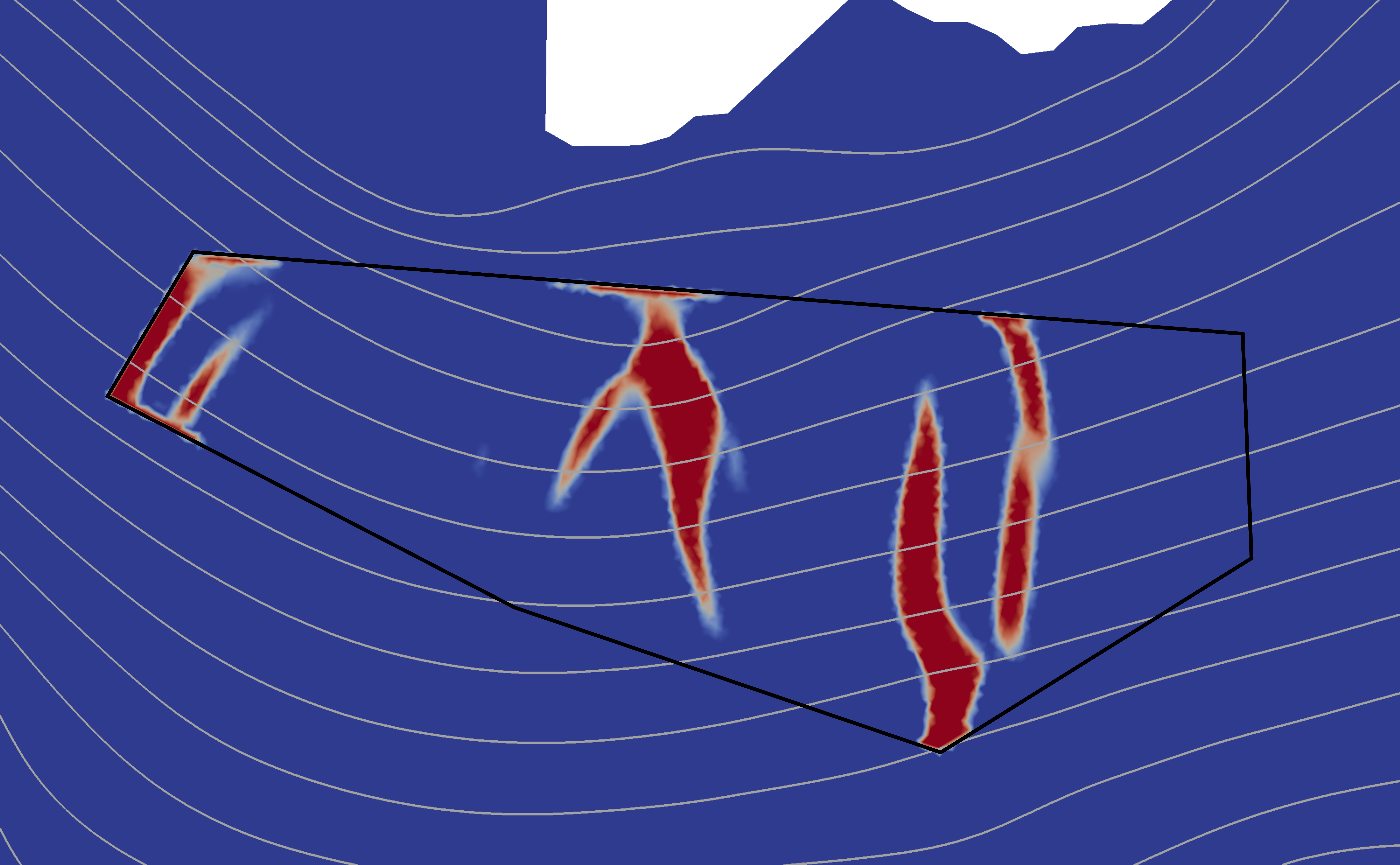}
    \caption{Farm 2 (251 turbines)}
    \label{fig:multi_farm_optimisation_farm_2_team}
  \end{subfigure}
  \begin{subfigure}[b]{0.48\textwidth}
    \includegraphics[width=\textwidth]{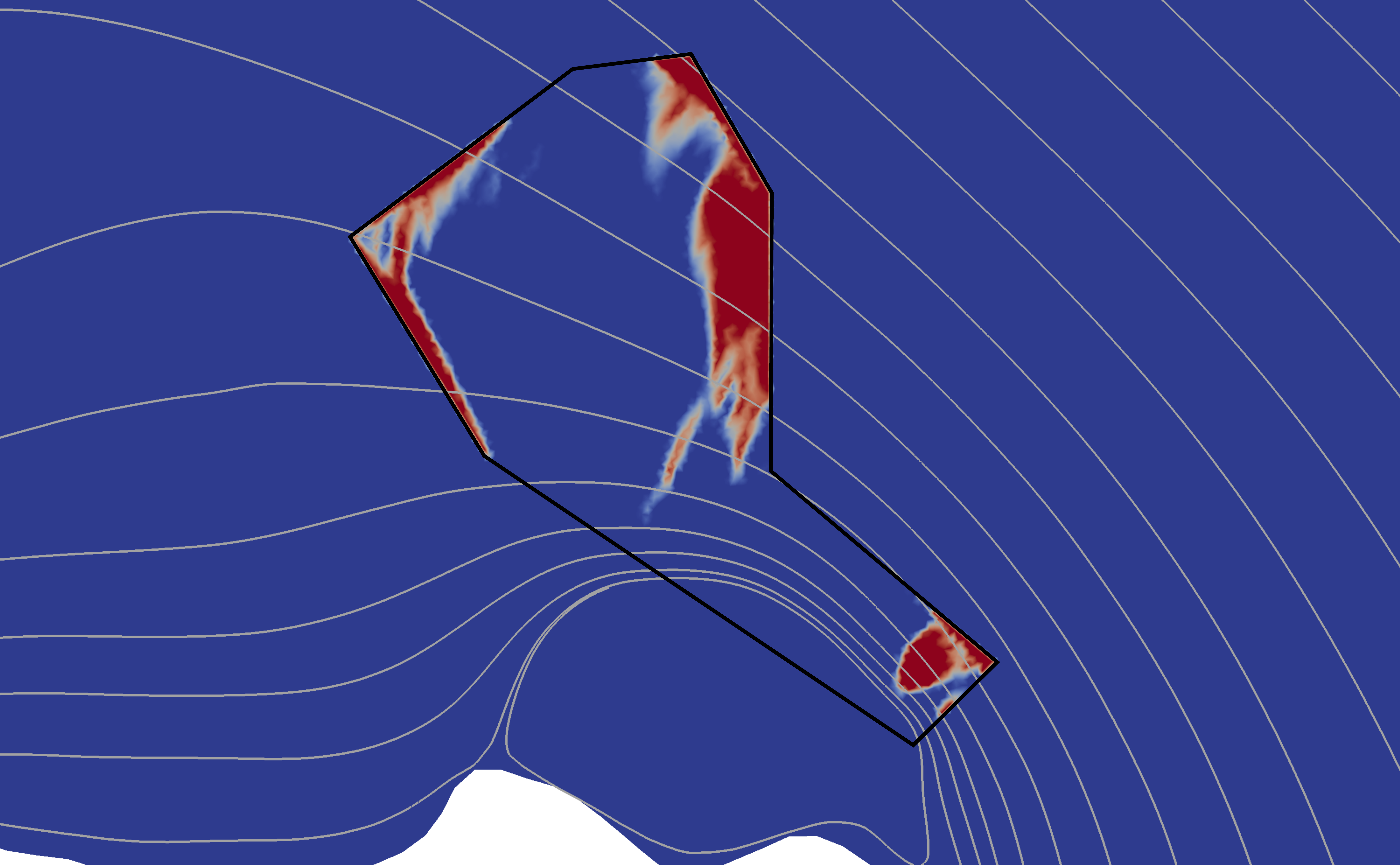}
    \caption{Farm 3 (249 turbines)}
    \label{fig:multi_farm_optimisation_farm_3_greedy}
  \end{subfigure}
  \begin{subfigure}[b]{0.48\textwidth}
    \includegraphics[width=\textwidth]{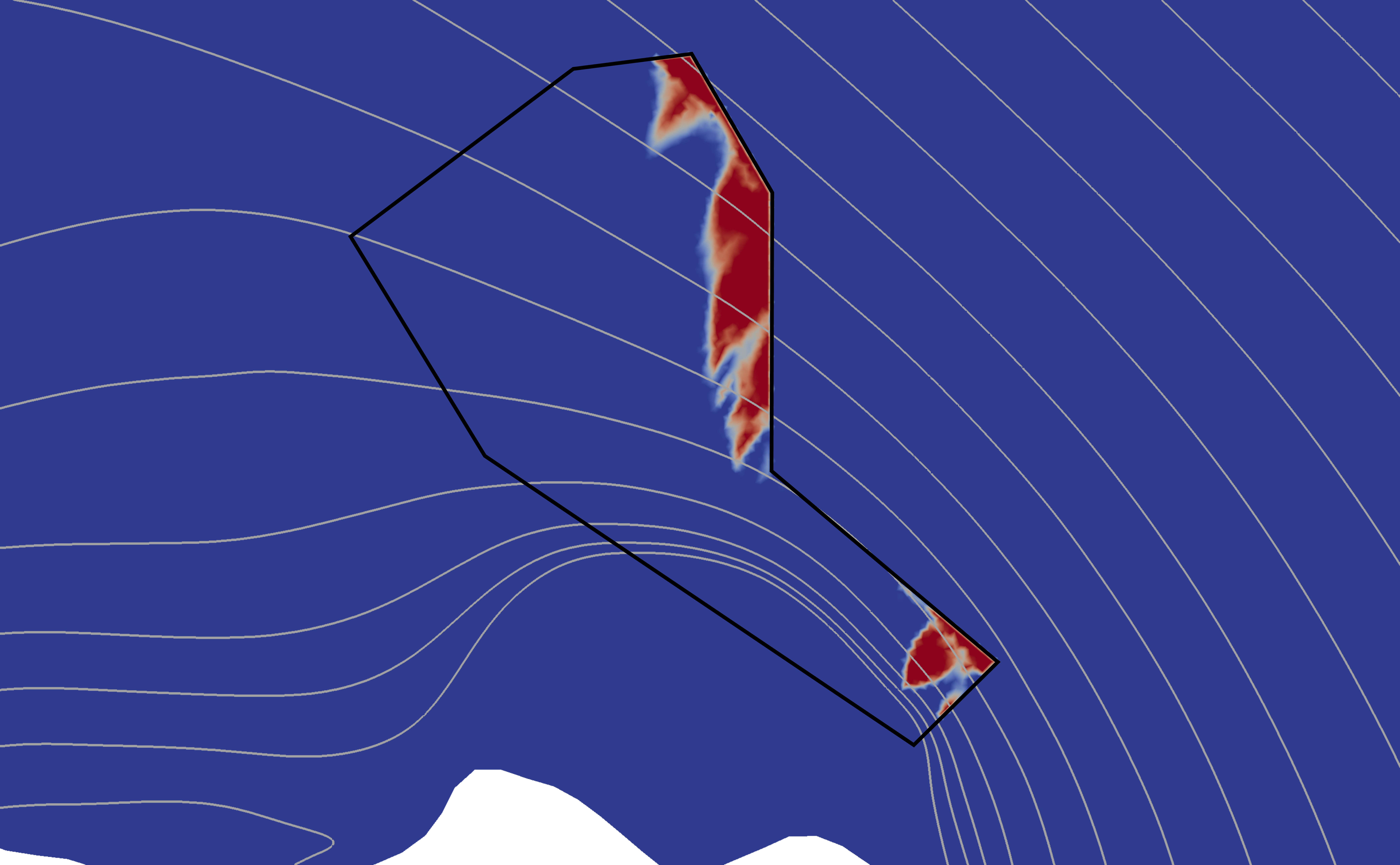}
    \caption{Farm 3 (162 turbines)}
    \label{fig:multi_farm_optimisation_farm_3_team}
  \end{subfigure}
  \begin{subfigure}[b]{0.48\textwidth}
    \includegraphics[width=\textwidth]{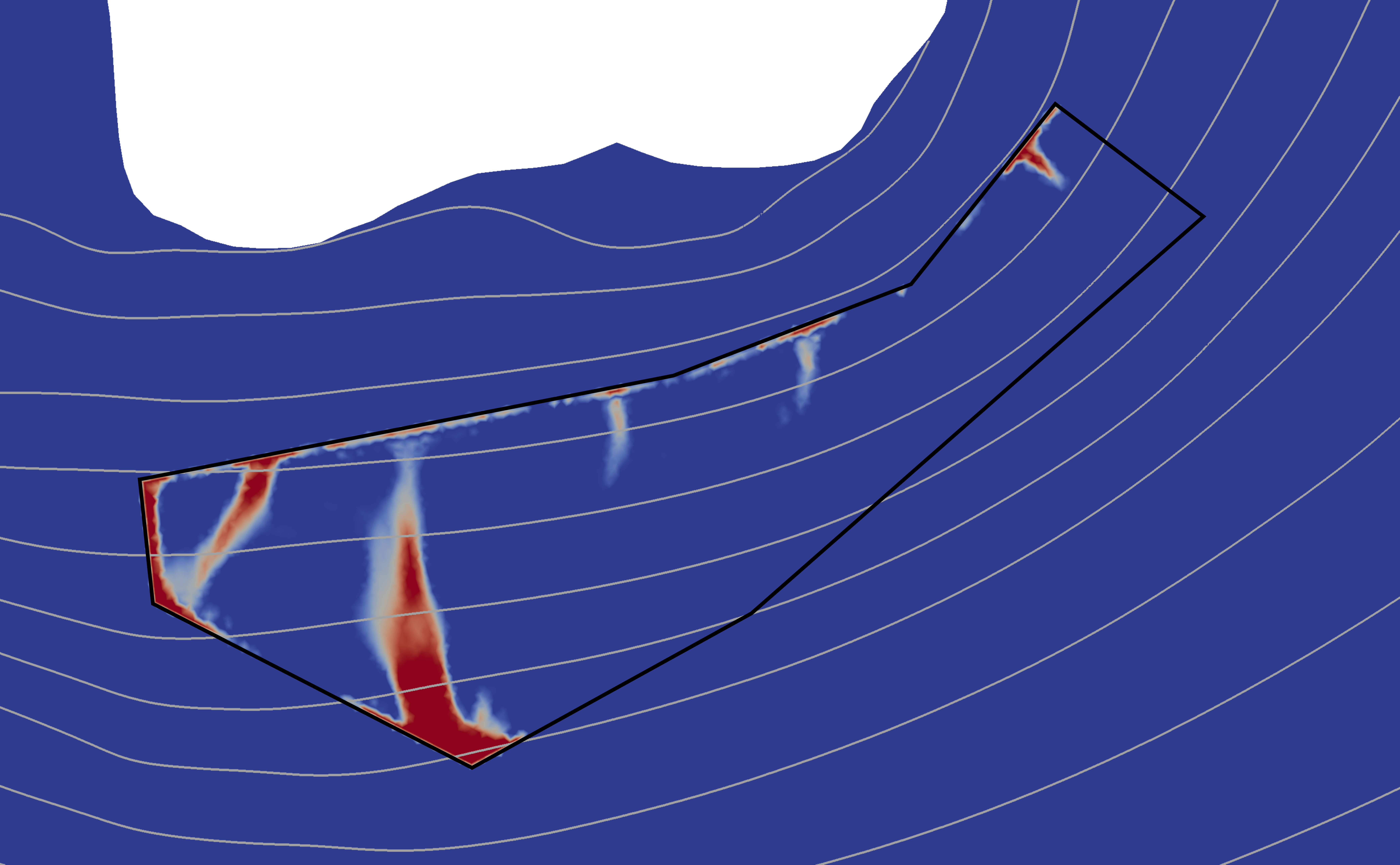}
    \caption{Farm 4 (156 turbines)}
    \label{fig:multi_farm_optimisation_farm_4_greedy}
  \end{subfigure}
  \begin{subfigure}[b]{0.48\textwidth}
    \includegraphics[width=\textwidth]{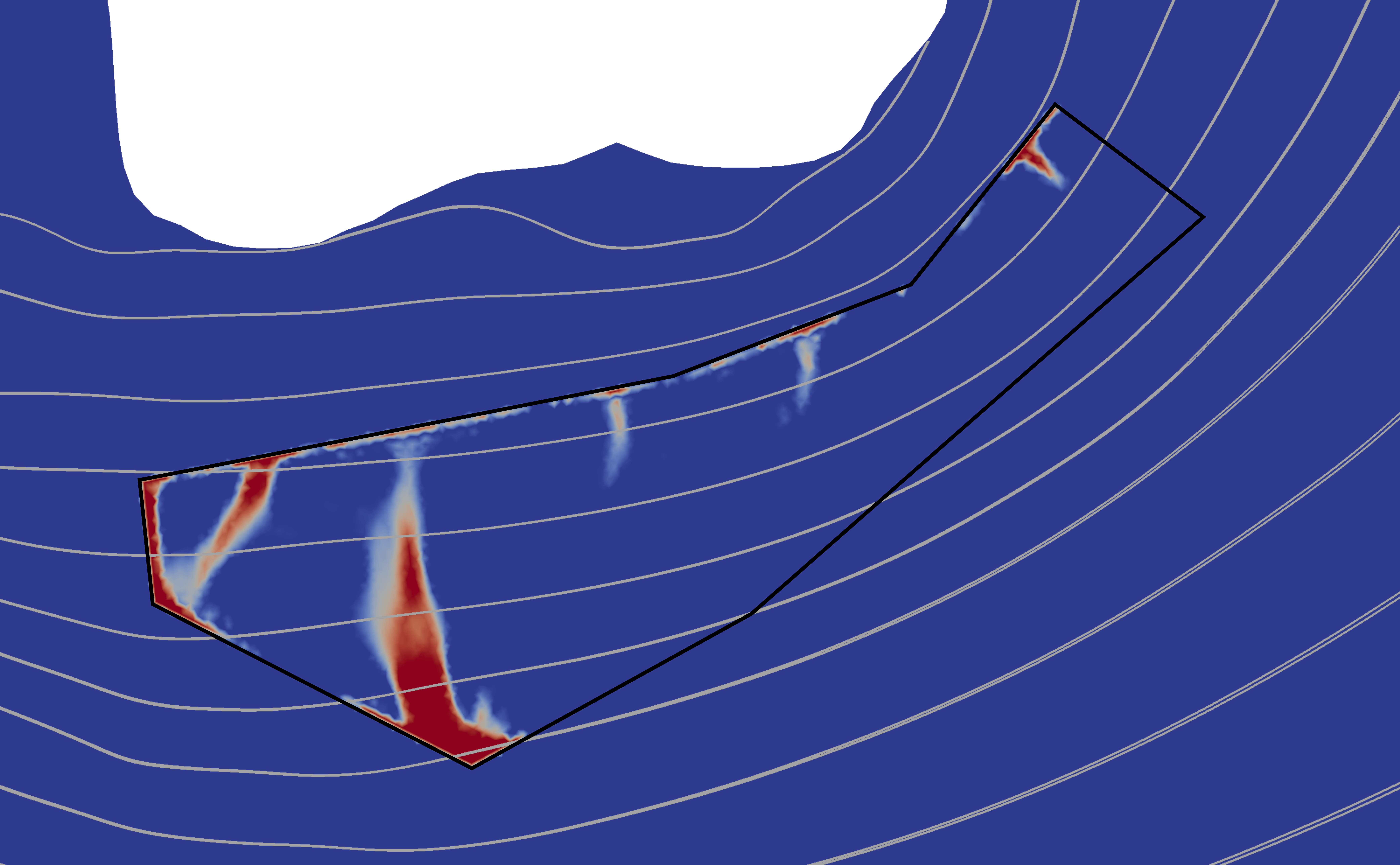}
    \caption{Farm 4 (153 turbines)}
    \label{fig:multi_farm_optimisation_farm_4_team}
  \end{subfigure}
  \caption{The optimal turbine densities for the four farms in the Pentland Firth.
          The designs on the left were obtained by individually optimising each
          farm, not taking into account the existence of the other farms,
          while the designs on the right were obtained from a multi-farm
          optimisation to maximise the overall profit of all farms.}
  \label{fig:multi_farm_optimisation_results_layouts}
\end{figure}

The optimal design of the multi-farm optimisation is shown in
figure~\ref{fig:multi_farm_optimisation_results_layouts} (right column).
The average profit, power, cost and optimal
number of turbines are listed in table~\ref{tab:multi_farm_results} (values in
brackets). When comparing these numbers with results form the individual
optimisations (non-bracketed values in table~\ref{tab:multi_farm_results}) it
becomes clear
that the individual design optimisations consistently overestimated each farm's
potential.
Compared to the designs from the individual optimisations, the differences are
marginal in farm 1 and 4. This can be explained from the fact that these are
fairly far apart and not directly in each others wake. The interaction between
farm 2 and 3 is stronger however, due to their close proximity and alignment
directly in each other's wake. The fact that farm 3 is more affected than farm 2 when taking
other farms into account in the optimisation can be ascribed to differences in
geometry. The flow through farm 2 is much more constrained due to the presence
of the island of Stroma, whereas in the case of farm 3 additional turbines can
easily lead to a redirection of the flow northward of the farm which would also
redirect it around Stroma island and thus farm 2.

In summary, the influence of multiple farms on one another can be significant, in particular if two
farms are in close proximity and share the same tidal stream. This
changes both the predicted profit and power production, as well as the optimal
tidal farm design.  The enhanced computational efficiency of the continuous
turbine approach presented in this work facilitates its use in the study of
fully coupled multi-farm optimisation.

\begin{table}
    \centering
    \begin{tabular}{l|cccccc}
      Site     & Site area &  Avg profit & Avg power & Cost & Number of \\
               & (km$^2$)  &  (MW)   & (MW)  & (MW) & turbines\\
      \hline
      Farm 1   & 9.9  & 31.2 (29.1) & 134.4 (130.4) & 103.2 (101.3) & 228 (224) \\
      Farm 2   & 3.1  & 195.7 (180.3) & 316.0 (293.9) & 120.3 (113.6) & 266 (251)\\
      Farm 3   & 2.3  & 70.5 (48.0) & 183.3 (121.1) & 112.8 (73.2) & 249 (162) \\
      Farm 4   & 2.8  & 65.5 (63.8) & 136.9 (133.2) & 70.4 (69.4) & 156 (153) \\
      Total      & 18.3 & 362.9 (321.2) & 769.6 (678.6) & 406.7 (357.5) & 899 (790) \\
    \end{tabular}
    \caption{Farm design optimisations results for four tidal farms in the
        Pentland Firth. The non-bracketed values are the results when each farm
        was individually optimised to maximise profit, neglecting the existence
        of the other farms. The values in brackets show the results when
        simultaneously optimising all four farms to maximise the overall profit.
        The discrepancy between the values indicates the risk of tidal farm
        developers to overpredict the farm potential when not taking into
        account other tidal farms.}
    \label{tab:multi_farm_results}
\end{table}

\section{Farm optimisation with complex constraints}\label{sec:complex}

\begin{figure}[t]
  \centering
  \begin{subfigure}[b]{0.48\textwidth}
    \includegraphics[width=\textwidth]{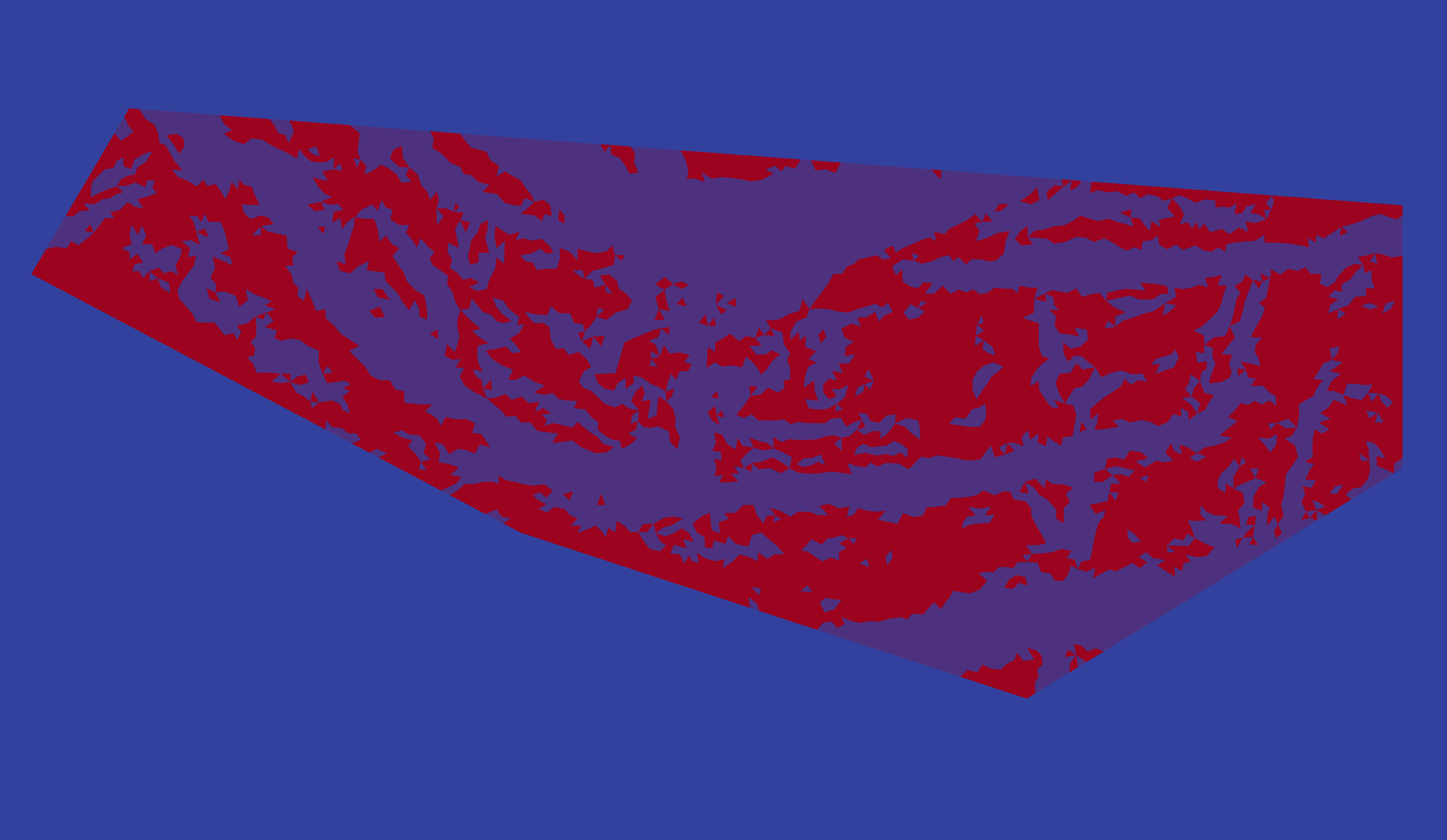}
    \caption{Original (shaded red) and masked (red) turbine farm area}
    \label{fig:single_farm_optimisation_results_farm_domain}
  \end{subfigure}
  \begin{subfigure}[b]{0.48\textwidth}
    \includegraphics[width=\textwidth]{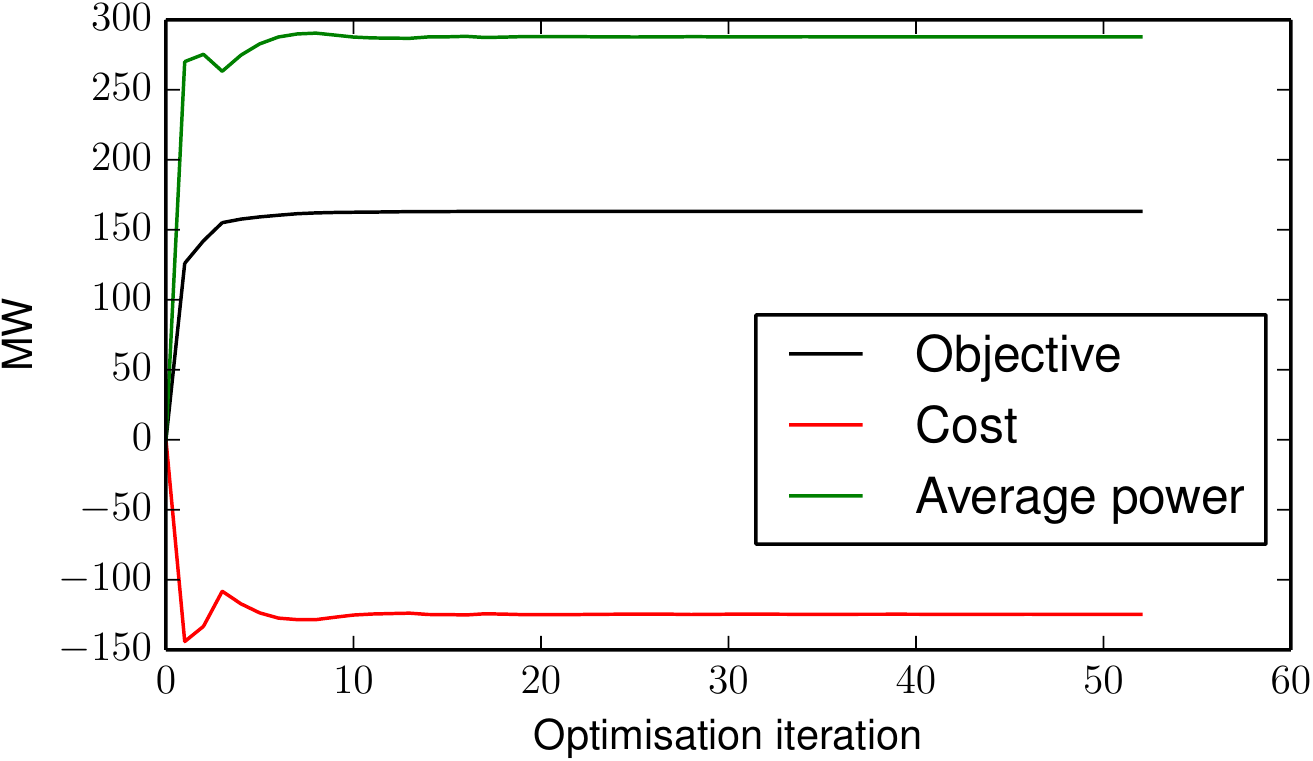}
    \caption{Optimisation iterations}
    \label{fig:single_farm_optimisation_results_farm_iterplot}
  \end{subfigure}
  \begin{subfigure}[b]{0.48\textwidth}
    \includegraphics[width=\textwidth]{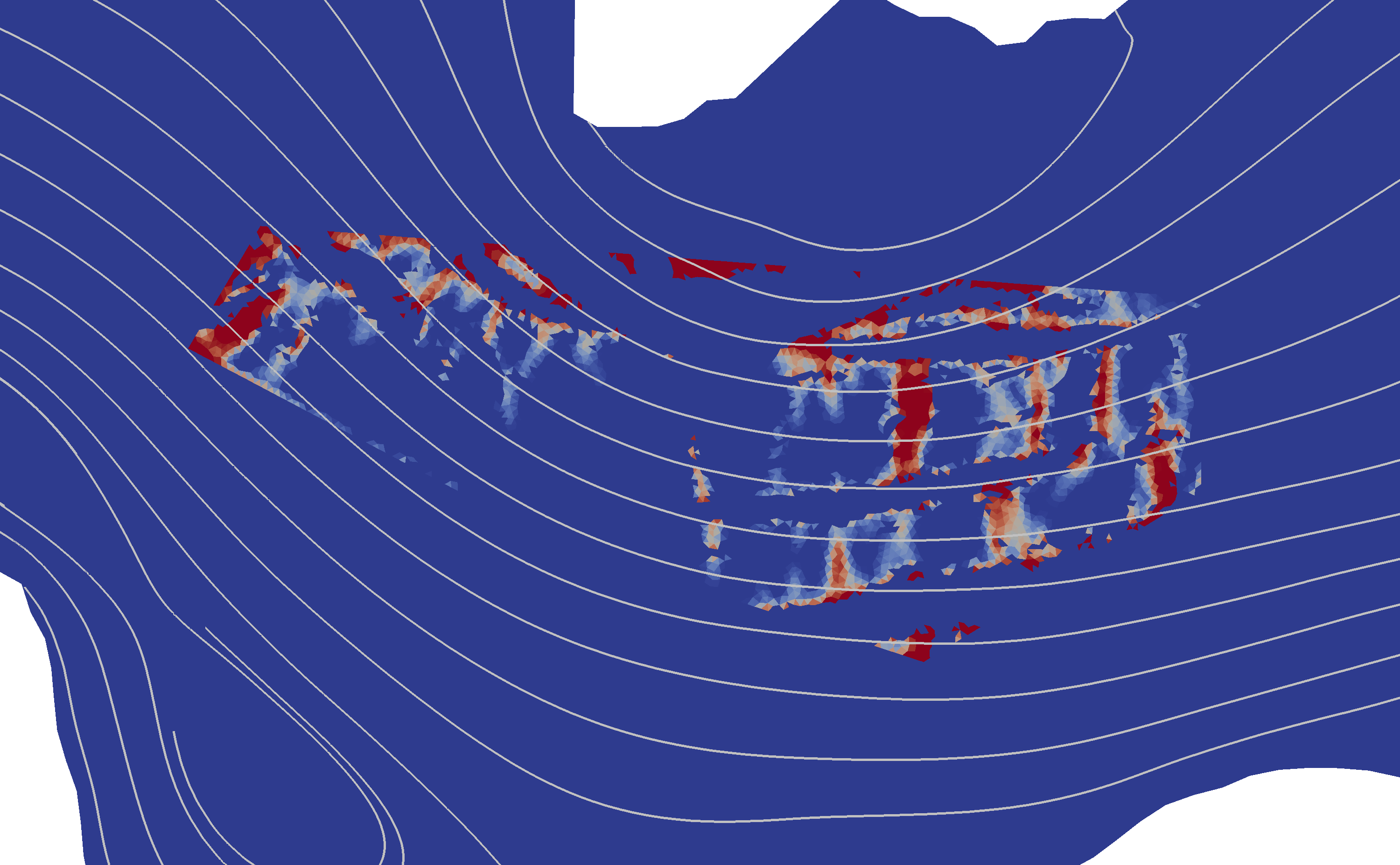}
    \caption{Optimal turbine density with stream lines during flood}
    \label{fig:single_farm_optimisation_results_farm_flood}
  \end{subfigure}
  \begin{subfigure}[b]{0.48\textwidth}
    \includegraphics[width=\textwidth]{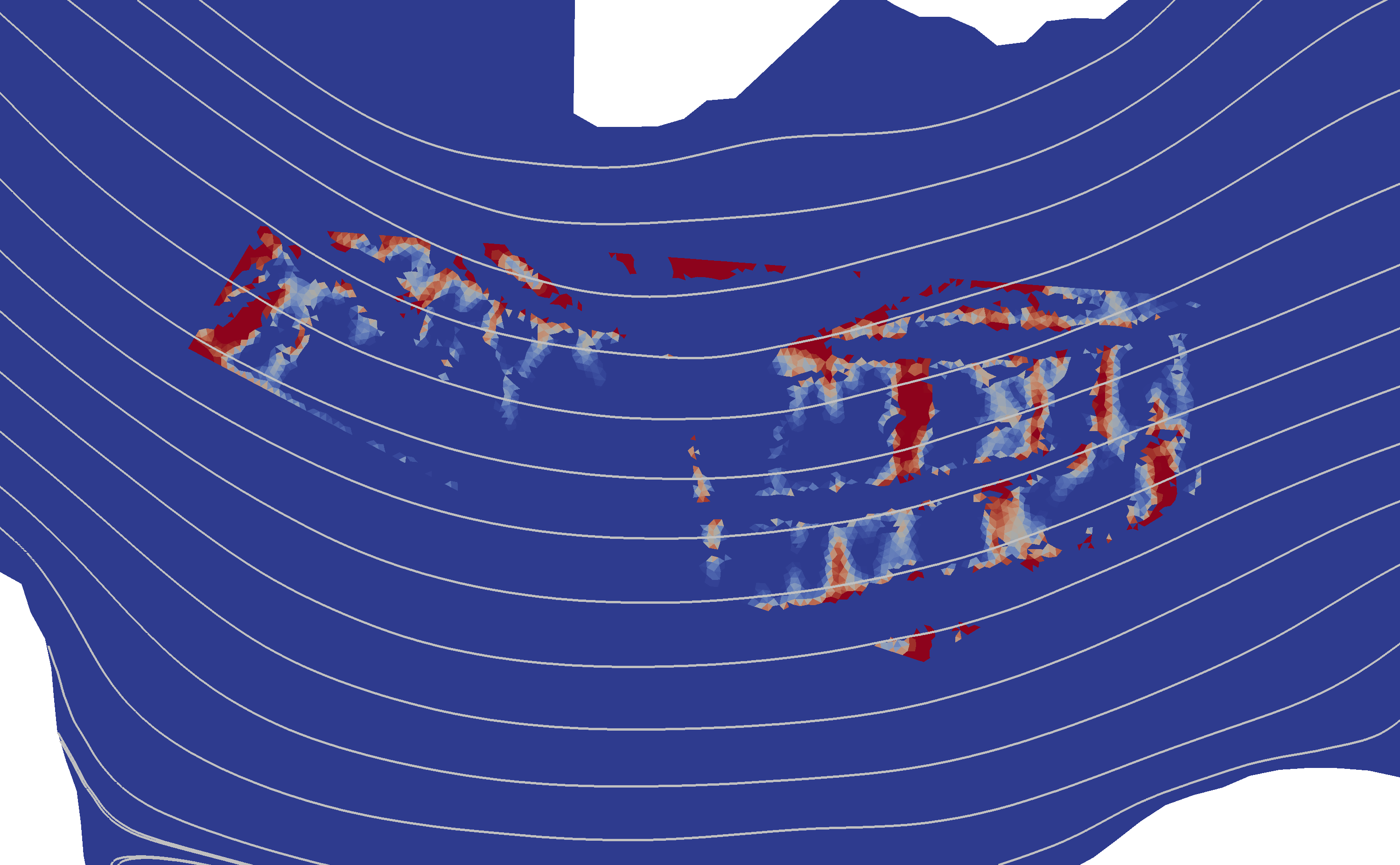}
    \caption{Optimal turbine density with stream lines during ebb}
    \label{fig:single_farm_optimisation_results_farm_ebb}
  \end{subfigure}
  \caption{The results of the farm optimisation with complex constraints}
  \label{fig:single_farm_optimisation_results}
\end{figure}

The practical design of a tidal farm is constrained by various restrictions. For
example, turbines may not be able to be placed at locations where the bathymetry
has a high slope, as well as in water that is either too shallow
or too deep. These sorts of restrictions must be considered when identifying suitable
areas for turbine installation.
When masking out the site areas which do not adhere to these constraints, one obtains a map
where turbines may be placed. Figure
\ref{fig:single_farm_optimisation_results_farm_domain} shows such a map for
farm 2 in the Inner Sound, with a maximum steepness condition $\|\nabla h\|
< 0.015$.  The map consists of many disjoint patches of feasible turbine
locations.  In the discrete farm design approach~\citep{funke2014}, optimising
such a site would be extremely challenging since the control space is not convex.
The continuous turbine approach however handles this case naturally.

To demonstrate this, we solved the continuous farm optimisation problem on the
masked farm area shown in
\ref{fig:single_farm_optimisation_results_farm_domain}. Otherwise, the problem setup
was equivalent to that in section \ref{sec:pentland-firth-optimisation}.

The convergence of the optimisation algorithm is shown in figure
\ref{fig:single_farm_optimisation_results_farm_iterplot}. It demonstrates the
efficiency of the approach: after $10$ iterations, the change in the goal quantity
per iteration has dropped below $3\times 10^{-7}$.  
Overall, the
optimisation was run for $54$ shallow-water solves and $52$ adjoint shallow-water
solves, but clearly could have been stopped earlier.

\begin{figure}[t]
  \centering
  \begin{subfigure}[b]{0.48\textwidth}
    \includegraphics[width=\textwidth]{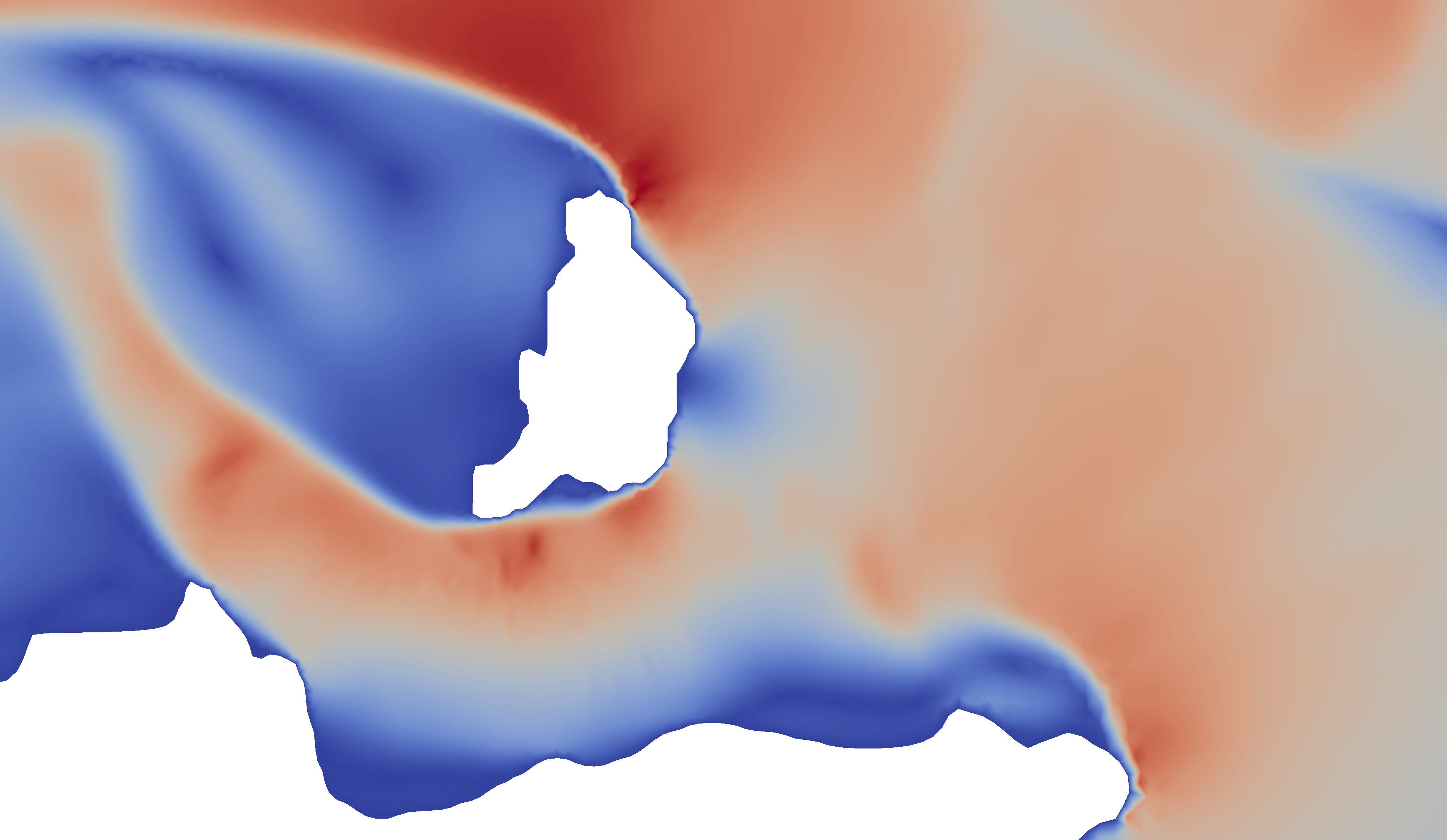}
    \caption{Flow speed without farm during ebb}
    \label{fig:single_farm_optimisation_flow_ebb_no_farm}
  \end{subfigure}
  \begin{subfigure}[b]{0.48\textwidth}
    \includegraphics[width=\textwidth]{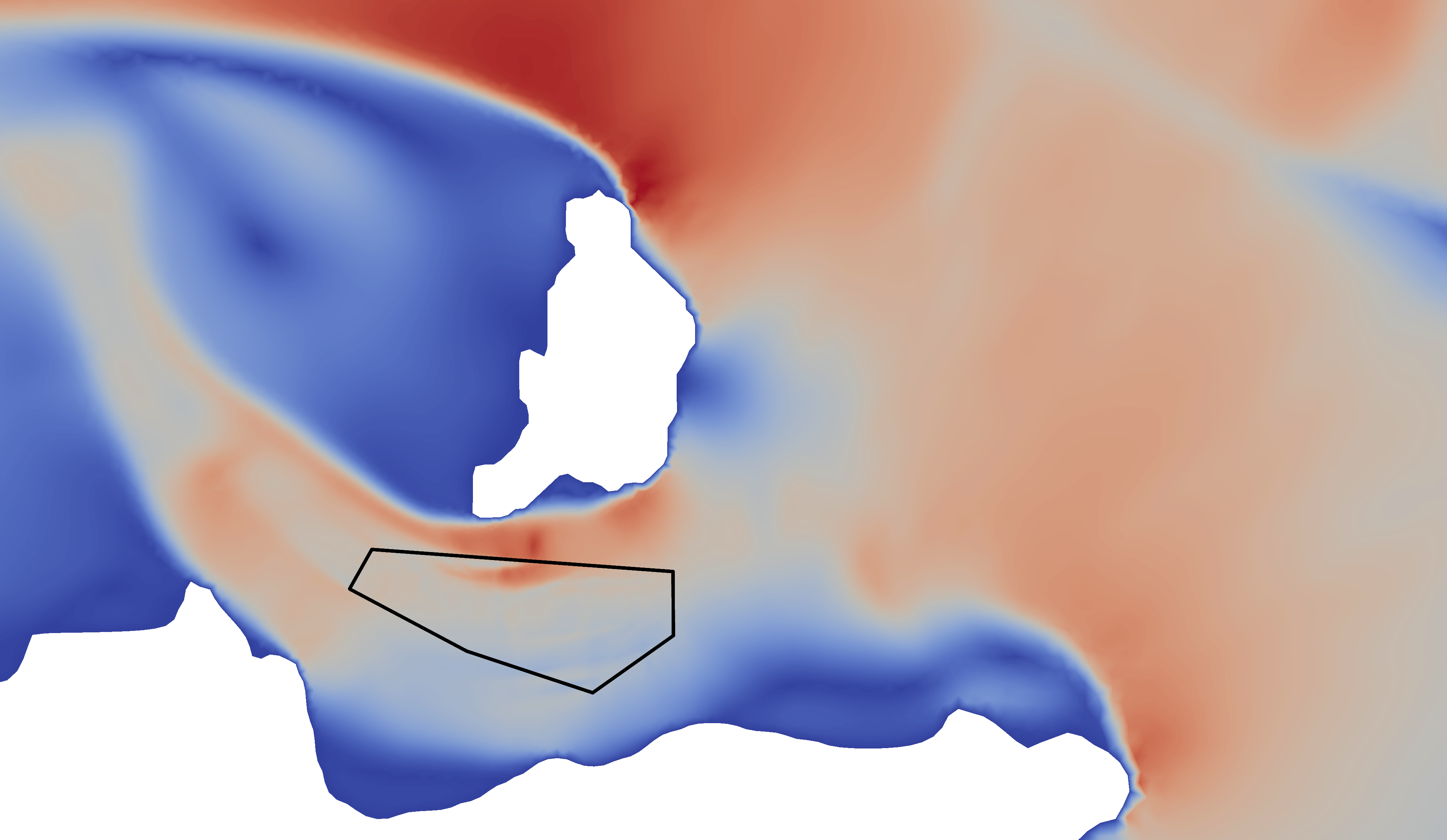}
    \caption{Flow speed with farm during ebb}
    \label{fig:single_farm_optimisation_flow_ebb_with_farm}
  \end{subfigure}
  \begin{subfigure}[b]{0.48\textwidth}
    \includegraphics[width=\textwidth]{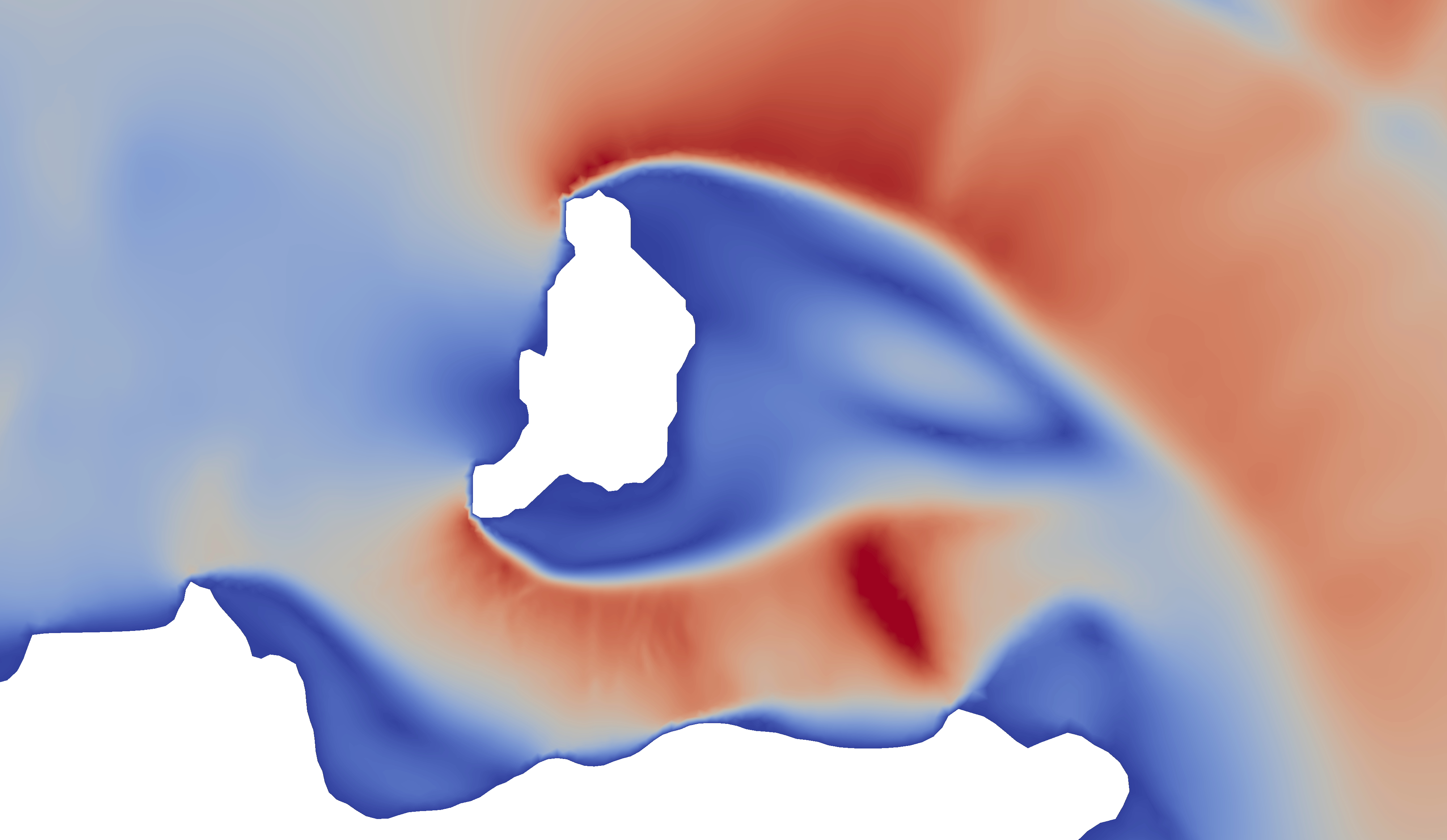}
    \caption{Flow speed without farm during flood}
    \label{fig:single_farm_optimisation_flow_flood_no_farm}
  \end{subfigure}
  \begin{subfigure}[b]{0.48\textwidth}
    \includegraphics[width=\textwidth]{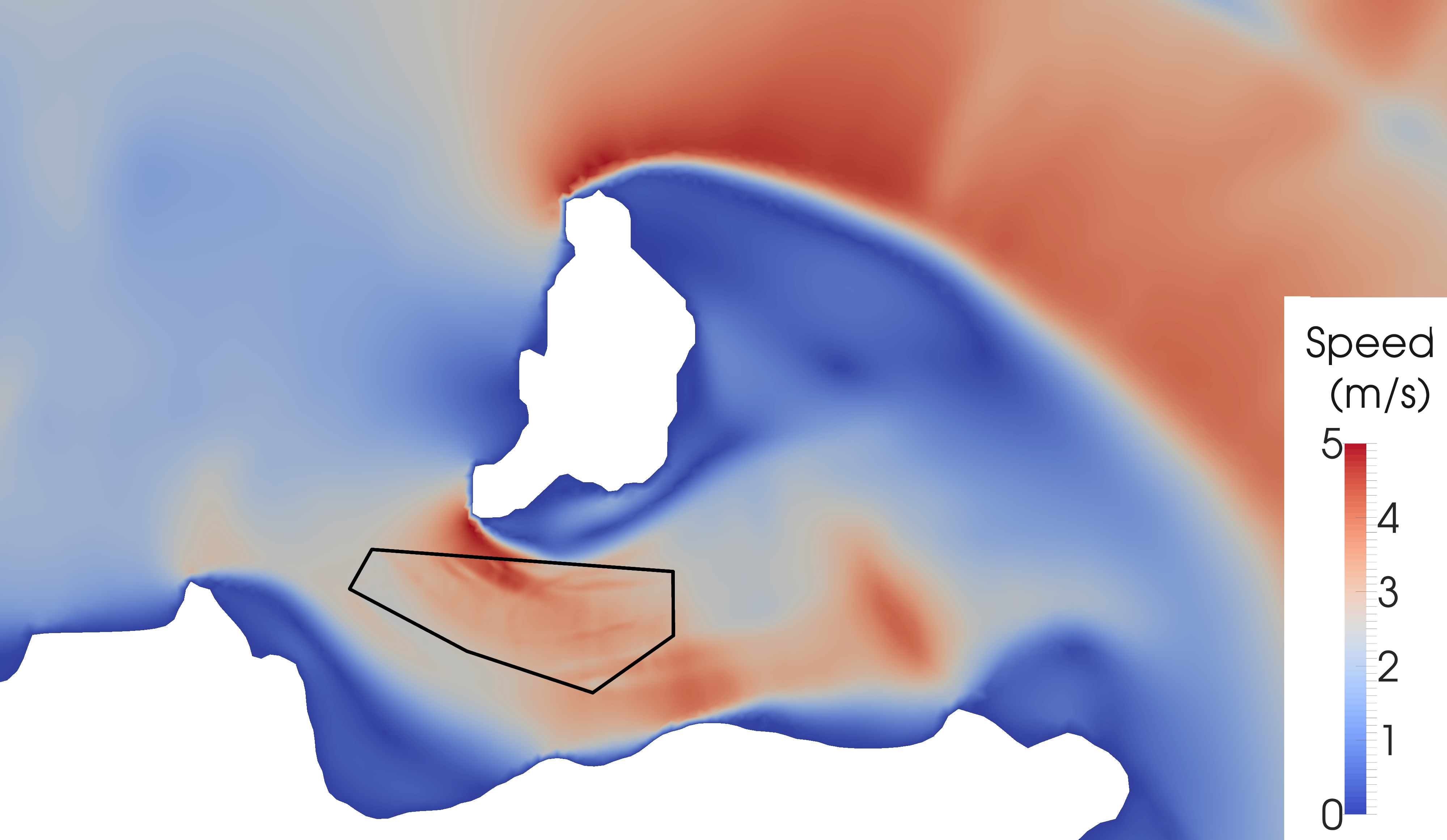}
    \caption{Flow speed with farm during flood}
    \label{fig:single_farm_optimisation_flow_flood_with_farm}
  \end{subfigure}
  \caption{The results of the farm optimisation with complex constraints}
  \label{fig:single_farm_optimisation_results_flow}
\end{figure}

Finally, figure \ref{fig:single_farm_optimisation_results_flow} visualises the flow speed
before and after the farm installation.
The optimal turbine density is plotted in
figures~\ref{fig:single_farm_optimisation_results_farm_flood} and
\ref{fig:single_farm_optimisation_results_farm_ebb}.
The optimisation algorithm positioned the areas of high turbine density (red)
near the farm boundaries and as barrages across two boundaries, with
much of the friction at the east and west sides of the farm and few turbines
in the farm interior.
The streamlines displayed in these plots show that these barrages are aligned
perpendicular to the flow. This has already been observed to be a characteristic of the
optimal designs obtained in
\citet{funke2014}. Also note that the constraints resulted in a
significantly different design with finer structures compared to the non-constrained
optimisation (figure~\ref{fig:multi_farm_optimisation_farm_2_greedy}). The
average power extraction of the optimal farm is $287.9$ MW.  The cost of the
farm (measured in power) is $124.6$ MW. The goal quantity (the profit
generating power extraction) is $163.3$ MW. The estimated number of turbines
was $275$, which corresponds to an average power extraction of $1.05$ MW per
turbine. Comparing these values with the values from the unrestricted
optimisation (table~\ref{tab:multi_farm_results}) shows that the restrictions led to
a slight negative impact in the profit and power production of the farm.

\section{Conclusions}

We formulated and solved the problem of finding the most profitable design of
tidal stream turbine farms as a mathematical optimisation problem constrained
by the shallow water equations.  We introduced a continuous approach that
describes the farm design using a turbine density function, and
demonstrated that this approach is consistent with an earlier approach where turbines are resolved
explicitly. The key benefits of this new continuous approach are its computational efficiency,
its flexibility and the robustness with respect to incorporating complex farm constraints.

The proposed approach can be used to answer a variety of questions: What is the
theoretically maximum extractable power of a tidal site? Taking into account installation
costs, what is the maximum overall profit of a farm?
To achieve this profit, what are the optimal number of turbines, and their positioning within the farm?
How do multiple tidal turbine farms impact upon each other?

So far, the proposed methodology has only been applied to the depth-averaged shallow water
equations, but it generalises also to fully three-dimensional models.
Considering the limitations of a depth-averaged model in predicting tidal
streams and in assessing tidal stream energy, future work should apply the same methodology
in a three-dimensional setting.

\section*{Acknowledgements}
This work was supported by the Research Council of Norway through a Centres of
Excellence grant to the Center for Biomedical Computing at Simula Research
Laboratory (project number 179578) and the UK's Engineering and Physical Science
Research Council (projects: EP/J010065/1 and EP/M011054/1).
Computations were performed on the Abel supercomputing cluster at the University of Oslo via
 NOTUR project NN9279K and NN9316K.
The authors would like to thank M. Nordaas for  valuable discussions.
We would like to
thank MeyGen Limited for supplying high resolution bathymetry data and ADCP data
in the Inner Sound area.

\bibliographystyle{elsarticle-harv}
\bibliography{literature}
\end{document}